\documentclass[eqsecnum,showpacs,aps,epsf,elsart]{revtex4}
\usepackage{graphicx,amssymb,bm,latexsym}
\usepackage{amsmath}
\begin{document}

\title{Stationary Spacetime from Intersecting M-branes}
\author{Kei-ichi Maeda$^{1,2,3}$
\footnote{e-mail address:maeda@gravity.phys.waseda.ac.jp}}
\author{Makoto Tanabe$^{1}$
\footnote{e-mail address:tanabe@gravity.phys.waseda.ac.jp}\\~}
\address{$^{1}$Department of Physics, Waseda University, 3-4-1 Okubo,
Shinjuku, Tokyo 169-8555, Japan}
\address{$^2$ Advanced Research Institute for Science and Engineering,
Waseda University, Shinjuku, Tokyo 169-8555, Japan}
\address{$^3$ Waseda Institute for Astrophysics, Waseda University,
Shinjuku,  Tokyo 169-8555, Japan}
\date{\today}

\begin{abstract}

 We study a stationary ``black" brane in M/superstring theory.
Assuming BPS-type relations between the first-order derivatives
of metric functions, we present general stationary black brane solutions 
 with a traveling wave for the Einstein equations in $D$-dimensions. 
The solutions are given by a few independent harmonic equations
(and plus the Poisson equation). General solutions are constructed 
 by superposition of a complete set of those harmonic functions.
Using the hyperspherical coordinate system,
we explicitly give the solutions in 11-dimensional
M theory for the case with M2$\perp$M5 intersecting branes
and a traveling wave.
Compactifying these solutions into five dimensions, we show that 
 these solutions include  the BMPV black hole 
and the Brinkmann wave solution.
We also find new solutions similar to the Brinkmann wave.
We prove that the solutions preserve the 1/8 supersymmetry 
if the gravi-electromagnetic field 
${\cal F}_{ij}$, which is a rotational part of gravity,
is self-dual. 
We also discuss 
non-spherical ``black" objects (e.g., a ring topology and 
 an elliptical shape)
by use of other curvilinear coordinates.

\end{abstract}

\pacs{ }

\maketitle

\section{Introduction}

Black holes are now one of the most important subjects in string theory. 
The Beckenstein-Hawking black hole entropy of an extreme black hole is
obtained in string theory 
by statistical counting of the corresponding microscopic 
states \cite{Strominger_Vafa}.
While, we have found several
 interesting black hole solutions in supergravity
theories \cite{Gibbons, CMP, Myers, Gibbons_Maeda, GHS, Horowitz_Strominger},
which are obtained as
an effective theory of a superstring model in a low energy limit.
We also know black hole solutions in a higher-dimensional 
spacetime \cite{Tangherlini, Myers_Perry},
which play a key role in a unified theory such as string theory.
In higher dimensions, because there is no uniqueness theorem of
black holes \cite{GIS, SUSYunique, Elvang_Emparan}, we have a variety of ``black" objects
such as a black brane \cite{Khuri_Myers, Duff_Lu, DLP, Callan_Maldacena}.
One of the most remarkable solutions is a black ring, which horizon 
has a topology of $S^1\times S^2$ \cite{Emparan_Reall}.

Among such ``black" objects, supersymmetric ones are very important.
The black hole solutions in a supergravity 
include the higher-order effects of a string coupling constant,
although these are solutions in a low energy limit.
On the other hand, the counting of states of corresponding branes
is performed at the lowest order of a string coupling.
The results of these two calculations need not coincide each other.
However, if there is supersymmetry,
these should be the same because the numbers of dynamical freedom 
cannot be different in these BPS representations.
Therefore, supersymmetric black hole (or black ring) solutions are often 
 discussed in many literature 
 \cite{Kallosh, FKS, Cvetic_Youm, Cvetic_Tseytlin, BCWKLM}.

The classification of supersymmetric solutions in minimal ${\cal N}=2$ 
supergravity in $D=4$ was first performed by a time-like or null 
Killing spinor \cite{tod}. Recently, solutions 
 in minimal ${\cal N}=1$ supergravity in $D=5$ have been classified into 
two classes 
by use of G-structures analysis \cite{sabra, GMT, GGHPR, gauntlett, Gauntlett_Pakis}. 
The six-dimensional minimal supergravity has also 
been discussed \cite{gutowski6d}.

However, the fundamental theory is constructed in either ten or eleven 
 dimensions.
When we discuss the entropy of black holes, 
we have to show the relation between those supersymmetric black holes
and more fundamental ``black" branes in either $D=10$ or 11,
from which we obtain  ``black" holes (or rings) via compactification.
The entropy is microscopically 
described by the charges of branes \cite{Das, Cvetic_Hull}.
A supersymmetric 
 rotating solution is obtained by
compactification from M or type II supergravity
  \cite{BMPV,herdeiro}. 
The supersymmetric rotating black ring solution  is   found \cite{EEMR04, 
EEMR05_1, Bena_Kraus, Kraus_Larsen, EEMR05_2}. 
Such solutions are  
obtained also in lower dimensions. 
These solutions are in fact new classes of rotating solutions 
 in four- or five-dimensional supergravity. 
The existence of such solutions suggests that the uniqueness theorem
of black holes is no longer valid even in supersymmetric 
 spacetime if the dimension is 
five or higher \cite{gutowski}.  Thus we may need 
to construct more generic ``black" brane solutions 
in the fundamental theory
and the black holes
by some compactification.
 M-theory is the best candidate for such a unified theory.
 Since its low energy limit coincides with the 
 eleven-dimensional supergravity, it provides a natural 
framework to study  ``black" brane or BPS brane solutions.

 In this paper we study a class of intersecting brane 
 solutions in 
 $D$-dimensions with a  $(d-1)$-dimensional transverse conformally
flat space. 
 We start with a generic form of the metric and solve the 
 field equations of the supergravity (the Einstein equations and
the equations for form fields). Assuming  the intersection 
 rule for the intersecting branes, which is the same as that derived in
a spherically symmetric case \cite{Ohta, OPS, Miao_Ohta},
we derive the equations for each metric.
We find that most metric components are described by 
harmonic functions, which are independent.
One metric component $f$, which corresponds to a traveling wave,
is usually given by the Poisson equation, which source term is 
given by the quadratic form of the ``gravi-electromagnetic" field 
${\cal F}_{ij}$.
In some configuration of branes, e.g., for two intersecting
charged branes (M2$\perp$M5), the source term vanishes.
As a result, we find only independent harmonic functions.
Hence, we can easily construct arbitrary solution by superposing 
those harmonics.
In order to preserve 1/8 supersymmetry, we have to impose
that ${\cal F}_{ij}$ is self-dual.

In \S II,
 we consider the  dilaton coupling gauged 
supergravity actions in $D$ dimensions, and 
derive the basic equations 
for a stationary ``black" brane, which  extra
$(D-d)$-dimensions are filled by several branes,  with a traveling wave. 
The solutions are given by 
harmonic functions on the $(d-1)$-dimensional Euclidian space. 
In \S III we 
construct the solution in eleven  (or ten ) dimensions. 
In \S IV, we present the explicit solutions
for $d=5$
by use of a hyperspherical coordinate system.
We recover the BMPV solution  \cite{BMPV,herdeiro} 
and the Brinkmann wave  \cite{Brinkmann}
as a special case.
The concluding remarks follow in
\S V.
In Appendix A,
we prove the 1/8
supersymmetry is preserved in our stationary ``black" brane solutions,
if ${\cal F}_{ij}$ is self-dual.
We also present some explicit solutions by use of different coordinate systems
(hyperelliptic and hyperpolorical coordinates)
in Appendix B.
 
\section{Basic Equations for  a Stationary Spacetime  with Branes}

We first present the basic equations for 
a stationary spacetime with intersecting branes
and describe 
how to construct generic solutions.
We consider the
following bosonic sector of a low energy 
effective action of superstring theory
or M-theory 
in $D$ dimensions $(D \leq 11)$:
\begin{eqnarray}
S=\frac{1}{16\pi G_D}\int d^DX\sqrt{-g}
\left[{\cal R}-\frac{1}{2}(\nabla\varphi)^2
-\sum_A\frac{1}{2\cdot n_A!}
e^{a_A\varphi}F_{{\bf n_A}}^2\right]
\,,
\label{action0}
\end{eqnarray}
where ${\cal R}$ is the Ricci scalar of a spacetime metric $g_{\mu\nu},
F_{{\bf n_A}}$ is the field strength  of an arbitrary form with a degree
$n_A (\leq D/2)$, and
$a_A$  is its coupling constant with a dilaton field $\varphi$. 
Each  index $A$ describes a different type  
 of brane. 
 Although we leave the
spacetime dimension $D$ free, the present 
action is most suitable for  describing the bosonic part of $D=10$ or
$D=11$ supergravity.

The equations of motion are written in the following forms: 
\begin{eqnarray}
&&{\cal R}_{\mu\nu}=\frac{1}{2}\partial _\mu\varphi\partial _\nu
\varphi+\sum_A\Theta_{{\bf n_A}\mu\nu},\nonumber\\
&&\nabla^2\varphi =\sum_A\frac{a_A}{2\cdot 
n_A!}e^{a_A\varphi}F_{{\bf n_A}}^2,\nonumber\\
&&\partial _{\mu _1}(\sqrt{-g} ~e^{a_A\varphi}F_{{\bf n_A}}^{~\mu _1
\cdots\mu_{n_A}})=0
\,,
\label{eqn:eom}
\end{eqnarray}
where $\Theta_{{\bf n_A}\mu\nu}$ is the 
stress-energy tensor of the $n_A$-form, which is given by
\begin{eqnarray}
\Theta_{{\bf n_A}\mu\nu}=\frac{1}{2\cdot n_A!}
e^{a_A\varphi}\left[n_A{F_{{\bf n_A} \mu}}^{\rho\cdots
\sigma}F_{{\bf n_A} \nu\rho\cdots\sigma}-\frac{n_A-1}{D-2}
F_{{\bf n_A}}^2g_{\mu\nu}\right]
\,.
\end{eqnarray}
We also have an additional equation, which is  the Bianchi
identity for the $n_A$-form, i.e.,
\begin{eqnarray}
\partial_{[\mu}F_{{\bf n_A}\mu _1\cdots\mu _{n_A}]}=0
\,.
\label{eqn:bianchi}
\end{eqnarray}
This is automatically satisfied if we introduce the potentials of 
$n_A$-form.

As for a metric form for a spacetime with intersecting branes, we assume 
the following metric form \cite{herdeiro}:
\begin{eqnarray}
ds^2&=&2\theta^{\hat{u}}\theta^{\hat{v}}
+\sum _{i=1}^{d-1}(\theta^{\hat{i}})^2
+\sum _{\alpha=2}^p(\theta^{\hat{\alpha}})^2
\,,
\end{eqnarray}
where $D=d+p$ and the dual basis $\theta^{\hat A}$ are given by 
\begin{eqnarray}
\theta^{\hat{u}}=e^\xi du, \quad \theta^{\hat{v}}=e^\xi
\left(dv+fdu+{{\cal A}\over \sqrt{2}}\right), 
\quad \theta^{\hat{i}}=e^\eta dx^i, \quad\theta^{\hat{\alpha}}
= e^{\zeta
_\alpha}dy^\alpha
\,.
\end{eqnarray}
Here we have used light-cone coordinates; 
$u=-(t-y_1)/\sqrt{2}$ and $v=(t+y_1)/\sqrt{2}$. 
This metric form includes rotation of spacetime and a traveling wave.
Since we are interested in a stationary solution,
we assume that the metric components 
$f$, ${\cal A}={\cal A}_i dx^i$, $\xi,\eta$
and
$\zeta_\alpha$ depend only on the spatial coordinates $x^i$
in $d$-dimensions, which coordinates are given by $\{t, x^i
(i=1, 2, \cdots, d-1)\}$. 
In this setting, we set  each 
brane $A$ in a submanifold of 
$p$-spatial dimensions, which coordinates are given by $\{y_\alpha
(\alpha=1, 2, \cdots, p)\}$.  
Note that the solution in this metric form 
is invariant under the
gauge transformation, 
${\cal A}\rightarrow {\cal A}+d\Lambda, \quad v\rightarrow v-\Lambda/\sqrt{2}$.

As for the $n_A$-form field with a $q_A$-brane, 
we assume that  the source brane exists in the coordinates
$\{y_1,y_{\alpha_2},\cdots, y_{\alpha_{q_A}}\}$.
The  form field generated by 
an ``electric" charge 
is given by the following form:
\begin{eqnarray}
F_{{\bf n_A}}&=&\partial _jE_{A} dx^j\wedge
du\wedge dv
\wedge dy_2\wedge\cdots\wedge dy_{q_A}
+{1\over \sqrt{2}}
\partial_i B^{A}_j dx^i\wedge dx^j\wedge du
\wedge dy_2\wedge\cdots\wedge dy_{q_A}
\label{e_field}
\,,
\end{eqnarray}
where  $n_A=q_A+2$ and $E_{A}$ and $B^{A}_j$ are scalar and vector 
potentials. 
This setting automatically 
guarantees the Bianchi identity  (\ref{eqn:bianchi}).

We can also discuss the  form field generated by 
a ``magnetic" charge   
by use of a dual ${}^*{n}_A$-field with
${}^*{q}_A$-brane, 
which is obtained  by 
a dual transformation of the $n_A$-field with a $q_A$-brane
 (${}^*{n}_A\equiv D-n_A, {}^*{q}_A\equiv {}^*{n}_A-2$).
In other words, the field components of $F_{{\bf n_A}}$ 
generated by a ``magnetic" charge
are described by the same form of (\ref{e_field}) of the dual field 
${}^*F_{{\bf {n}_A}}=F_{{\bf {}^*{n}_A}}$. We then 
treat $F_{{\bf {}^*{n}_A}}$,
which is generated by a ``magnetic" charge,
 as another independent form field with a
different brane
from $F_{{\bf n_A}}$,
which is generated by an ``electric" charge, 
when we sum up by the types of branes $A$.

Setting
\begin{eqnarray}
H_A&=&\exp\left[-\left(
2\xi+\sum_{\alpha=\alpha_2}^{\alpha_{q_A}}
\zeta_\alpha-{1\over 2}\epsilon_A
a_A\varphi
\right)\right]\\
V&=&\exp\left[
2\xi+(d-3)\eta+\sum_{\alpha=2}^{p}
\zeta_\alpha
\right]
\,,
\label{def:H_A}
\end{eqnarray}
where 
\begin{eqnarray}
\epsilon_A =\left\{\begin{array}{ll}
+1 & \mbox{$n_A$-form field ($F_{{\bf n_A}}$)} \\
-1 & \mbox{the dual field (${}^*F_{{\bf {n}_A}}$)}
\end{array}\right.
\,,
\end{eqnarray}
we find
the basic equations as follows:
\begin{eqnarray}
&&\partial^2f+\partial_j f\partial^j\ln V
={1\over 8}e^{2(\xi-\eta)}
\left[{\cal F}_{ij}^2
-{1\over 2}\sum_A 
\left({\cal F}_{ij}^{(A)}\right)^2
\right] 
\label{eqn:Ruu},
\\
&&\partial^2\xi+\partial_j\xi\partial^j\ln V
=\frac{1}{2(D-2)}\sum_A (D-q_A-3)H_A^{2}
(\partial E_A)^2
\label{eqn:Ruv},\\
&&\partial^j{\cal F}_{ij} +{\cal
F}_{ij}\partial^j\left[2\left(\xi-\eta\right)+
\ln V\right]
=\sum_A H_A
{\cal F}_{ij}^{(A)}
\partial^j
E_A
\label{eqn:Rui},\\
&&\left(\partial^2
\eta+\partial_l\eta\partial^l\ln V\right)\delta_i^j
+2\partial_i\xi\partial^j\xi+(d-3)\partial_i\eta\partial^j\eta
+\sum_{\alpha=2}^p
\partial_i\zeta_\alpha\partial^j\zeta_\alpha +
\partial_i\partial^j\ln
V-\left(\partial_i\eta\partial^j\ln V
+\partial^j\eta\partial_i\ln
V\right)
\nonumber\\
&&\quad=-\frac{1}{2} \partial_i\varphi\partial^j\varphi 
+\frac{1}{2}\sum_A H_A^{2}
\left[
\partial_i E_A \partial^j E_A-{q_A+1\over (D-2)}
 (\partial E_A)^2\delta_i^j \right]
\label{eqn:Rii},\\
&&\partial^2\zeta_\alpha+\partial_j\zeta_\alpha\partial^j
\ln V
=\frac{1}{2(D-2)}\sum_A\delta_{\alpha A} H_A^{2}
(\partial E_A)^2
\label{eqn:Raa},\\
&&\partial^2\varphi+\partial_j\varphi\partial^j
\ln V
=-\frac{1}{2}\sum_A\epsilon_A a_A H_A^{2}
(\partial E_A)^2
\label{eqn:phi},\\
&&\partial_j(H_A^2
V\partial^j E_A)=0,
\label{phi_A}\\
&&\partial^j\left(V
{\cal F}_{ij}^{(A)}\right)=0
\,,
\label{eqn:maxwell}
\end{eqnarray}
where 
$\partial_i$ is a partial derivative $\partial/\partial x^i$ in a
flat
$(d-1)$-space, $\partial^2\equiv \partial_i\partial^i$, and  ${\cal F}_{ij}$,
${\cal F}_{ij}^{(A)}$, and 
$\delta_{\alpha A}$ for each coordinate
$\alpha$  are defined by
\begin{eqnarray}
&&
{\cal F}_{ij}\equiv \partial_i {\cal A}_j-\partial_j {\cal A}_i
\nonumber \\
&&{\cal F}_{ij}^{(A)}\equiv 2H_A\left({\cal A}_{[i}\partial_{j]}H_A-
\partial_{[i}B_{j]}^A\right)
\nonumber \\
&&\delta_{\alpha A}\equiv \left\{\begin{array}{ll}
D-q_A-3 & ~~~~~\alpha ={\alpha_2,\ldots ,\alpha_{q_A}} \\
-(q_A+1) & ~~~~~\mbox{otherwise}
\end{array}\right.
\,.
\end{eqnarray}
The square bracket denotes the 
anti-symmetrization of indices, i.e., $X_{[i}Y_{j]}
\equiv {1\over 2}(X_iY_j-X_jY_i)$

Since $E_A$ appears just with a spatial derivative $\partial_i$,
we can replace it with $\tilde{E}_A=E_A-E_A^{(0)}$,
where $E_A^{(0)}$ is a constant, which is fixed by a boundary condition.
Using Eqs. (\ref{phi_A}) and (\ref{eqn:maxwell}), 
we obtain from Eqs.
(\ref{eqn:Ruv}), (\ref{eqn:Raa}) , (\ref{eqn:phi})  and (\ref{eqn:Rui}), 
\begin{eqnarray}
&&\partial^j \left[V\left(\partial_j \xi
-\frac{1}{2(D-2)}\sum_A(D-q_A-3)H_A^{2}
\tilde{E}_A\partial_j\tilde{E}_A \right)\right]=0
\label{el_xi},\\
&&\partial^j\left[V\left(\partial_j \zeta_\alpha
-\frac{1}{2(D-2)}\sum_A\delta_{\alpha A}H_A^{2}
\tilde{E}_A \partial_j \tilde{E}_A\right)\right]=0
\label{el_zeta},\\
&&\partial^j\left[V\left(\partial_j \varphi
+\frac{1}{2}\sum_A\epsilon_A a_A H_A^{2}
\tilde{E}_A  \partial_j\tilde{E}_A\right)\right]=0
\,.
\label{el_varphi}
\end{eqnarray}

This set of equations is a coupled system of elliptic-type differential
equations, for which it is very difficult to find general solutions.
Hence, in this paper, we assume
the following special relations:
\begin{eqnarray}
&&\partial_j \xi
=\frac{1}{2(D-2)}\sum_A(D-q_A-3)H_A^{2}
\tilde{E}_A\partial_j\tilde{E}_A 
\label{def:xi}
,\\
&&\partial_j \zeta_\alpha
=\frac{1}{2(D-2)}\sum_A \delta_{\alpha A}H_A^{2}
\tilde{E}_A \partial_j \tilde{E}_A
\label{def:zeta}
,\\
&&\partial_j \varphi
=-\frac{1}{2}\sum_A\epsilon_A a_A H_A^{2}
\tilde{E}_A  \partial_j\tilde{E}_A
\label{def:varphi}
\,,
\end{eqnarray}
which guarantee Eqs.
(\ref{el_xi}), (\ref{el_zeta})  and (\ref{el_varphi})
to be correct.

These equations
 are relations between the first-order derivatives
of variables  just as the BPS conditions.
The existence of supersymmetry
in the obtained solutions is shown in the Appendix.
Hence, 
these relations may be related to a BPS state, or 
an extremal ``black" brane solution in supergravity.
In fact, if ${\cal F}_{ij}^{(A)}$ is proportional
to ${\cal F}_{ij}$ and ${\cal F}_{ij}$ is self-dual,
we prove that 1/8 supersymmetry remains
in the solutions with M2$\perp$M5 branes
for $D=11$ supergravity theory.

$\eta$ is obtained from $\xi$, $\zeta_\alpha$ and $\varphi$
 as 
\begin{eqnarray}
\partial^j\eta&=&-\frac{1}{d-3}\partial^j\left(2\xi+\sum_{\alpha=2}^p
\zeta_\alpha-\ln V\right)\nonumber \\
&=&
-\frac{1}{2(D-2)}\sum_A (q_A+1)
H_A^{2}\tilde{E}_A\partial^j\tilde{E}_A+ \frac{1}{(d-3)}\partial^j \ln V 
\,.
\label{def:eta}
\end{eqnarray}
This gives
\begin{eqnarray}
\partial^2\eta+\partial_j\eta\partial^j\ln V&=&
{1\over V}\partial_j(V\partial^j \eta)\nonumber \\
&=&-\frac{1}{2V(D-2)}\sum_A (q_A+1)
\partial_j(H_A^{2}V\tilde{E}_A\partial^j\tilde{E}_A)
+ \frac{1}{(d-3)V}\partial^2  V\nonumber \\
&=&-\frac{1}{2(D-2)}\sum_A (q_A+1)
H_A^{2}(\partial\tilde{E}_A)^2
+ \frac{1}{(d-3)V}\partial^2  V
\,.
\label{ddeta}
\end{eqnarray}

We have, however, another equation for $\eta$, i.e., 
Eq. (\ref{eqn:Rii}),
which should be satisfied as well.
We have to find a solution which satisfies both
equations.
This consistency gives two conditions for $\tilde{E}_A$.
In order to derive them,
we first take a trace of Eq. (\ref{eqn:Rii}), which leads
to
\begin{eqnarray}
&&(d-1)(\partial^2
\eta+\partial_l\eta\partial^l\ln V)
+2(\partial\xi)^2+(d-3)(\partial\eta)^2
+\sum_{\alpha=2}^p (\partial\zeta_\alpha)^2 +
\partial^2\ln
V-2\partial_l\eta\partial^l\ln V
\nonumber\\
&&\quad=-\frac{1}{2} (\partial\varphi)^2
+\frac{1}{2(D-2)}\sum_A
[D-2-(q_A+1)(d-1)]H_A^{2}
 (\partial\tilde{E}_A)^2
\,.
\label{trace}
\end{eqnarray}

Substituting Eqs. (\ref{def:xi}), 
(\ref{def:zeta}), (\ref{def:varphi}) and 
(\ref{ddeta}) into Eq. (\ref{trace}), 
we find the first condition:
\begin{eqnarray}
\frac{1}{2} \sum_{A,B}M_{AB}H_A^{2} H_B^{2}\tilde{E}_{A}\tilde{E}_{B}
(\partial\tilde{E}_{A})(\partial\tilde{E}_{B})
-\sum_A  H_A^{2} (\partial\tilde{E}_A)^2
+4\left({d-2\over
d-3}\right)  V^{-1/2}\partial^2 V^{1/2}=0
\,,
\label{eqn:Rii2}
\end{eqnarray}
where
\begin{eqnarray}
M_{AB}={1\over (D-2)^2}\left[2(D-q_A-3)(D-q_B-3)+(d-3)(q_A+1)(q_B+1)
+\sum_{\alpha=2}^p \delta_{\alpha A}\delta_{\alpha B}\right] +{1\over
2}\epsilon_A
\epsilon_B a_A a_B
\,.
\label{eqn:MAB}
\end{eqnarray}

We have also a traceless part of Eq. (\ref{eqn:Rii}), 
which is written as
\begin{eqnarray}
&&2\left(\partial_i\xi\partial^j\xi-{1\over
d-1}(\partial\xi)^2\delta_i^j\right)+(d-3)\left(\partial_i\eta
\partial^j\eta-{1\over
d-1}(\partial\eta)^2\delta_i^j\right)
+\sum_{\alpha=2}^p \left(
\partial_i\zeta_\alpha\partial^j\zeta_\alpha-{1\over
d-1}(\partial\zeta_\alpha)^2\delta_i^j\right)
\nonumber \\
&&+{1\over 2}\left(\partial_i\varphi\partial^j\varphi-{1\over
d-1}(\partial\varphi)^2\delta_i^j\right)-{1\over
2}\sum_AH_A^2\left(\partial_i\tilde{E}_A\partial^j\tilde{E}_A-{1\over
d-1}(\partial\tilde{E}_A)^2\delta_i^j\right)
\nonumber \\
&&+\partial_i\partial^j \ln V-{1\over d-1}\partial^2\ln V\delta_i^j
-\left(\partial_i\eta\partial^j\ln V+\partial^j\eta\partial_i\ln
V -{2\over d-1}(\partial \eta)(\partial\ln V)\delta_i^j\right)=0
\,.
\label{traceless}
\end{eqnarray}

This equation gives the second condition:
\begin{eqnarray}
&&\frac{1}{2} \sum_{A,B}M_{AB}H_A^{2} H_B^{2}\tilde{E}_{A}\tilde{E}_{B}
(\partial_i\tilde{E}_{A})(\partial^j\tilde{E}_{B})
-\sum_A  H_A^{2} \partial_i\tilde{E}_A\partial^j\tilde{E}_A
\nonumber \\
&&
-{1\over d-1}\delta_i^j\left[\frac{1}{2} 
\sum_{A,B}M_{AB}H_A^{2} H_B^{2}\tilde{E}_{A}\tilde{E}_{B}
(\partial\tilde{E}_{A})(\partial\tilde{E}_{B})
-\sum_A  H_A^{2} (\partial\tilde{E}_A)^2
\right]
\nonumber \\
&&
-2(d-3)V^{1\over d-3}\left[
\partial_i\partial^j \left(V^{-{1\over d-3}}\right)
-{1\over d-1}\delta_i^j
\partial^2\left(V^{-{1\over d-3}}\right)\right]
=0
\,,
\label{traceless2}
\end{eqnarray}

We have to find a solution for two conditions (\ref{eqn:Rii2})
and (\ref{traceless2}).
Here we shall assume $V$= constant.
We shall also impose the condition $M_{AB}=0$ for
$A \neq B$, which is called 
the ``intersection" rule \cite{Ohta, OPS, Miao_Ohta}. 
This rule is derived in the case of ``spherically symmetric"
spacetime from the condition that each $E_A$ is independent. 
Our case is just an ansatz.

Suppose that the $q_A$-brane and $q_B$-brane are filled  in
different spatial dimensions, but those branes are crossing on
$\bar{q}_{AB}$ dimensions ($\bar{q}_{AB}<q_A, q_B$). Calculating 
(\ref{eqn:MAB}), we obtain
\begin{eqnarray}
M_{AB}=\bar{q}_{AB}+1-\frac{(q_A+1)(q_B+1)}{D-2}
+\frac{1}{2}\epsilon_A a_A\epsilon_B a_B
\,.
\end{eqnarray}
Since we assume that it vanishes for $A\neq B$,
 we obtain the crossing dimensions
$\bar{q}_{AB}$  as 
\begin{eqnarray}
\bar{q}_{AB}=\frac{(q_A+1)(q_B+1)}{D-2}-1
-\frac{1}{2}\epsilon_A a_A\epsilon_B a_B
\,.
\end{eqnarray}

Eqs. (\ref{eqn:Rii2})
and (\ref{traceless2}) are then reduced to
\begin{eqnarray}
&&\sum_{A}\left[\frac{1}{2} M_{AA}H_A^{2} \tilde{E}_A^2
- 1\right] 
H_A^{2}(\partial
\tilde{E}_A)^2=0\,,
\label{eqn:Rii3} \\
&&
\sum_{A}\left[\frac{1}{2} M_{AA}H_A^{2} \tilde{E}_A^2
- 1\right] 
H_A^{2}\left[\partial_i\tilde{E}_A\partial^j
\tilde{E}_A-{1\over d-1}(
\partial\tilde{E}_A)^2\delta_i^j\right]=0
\,.
\label{traceless3}
\end{eqnarray}
Hence, if  
\begin{eqnarray}
\frac{1}{2} M_{AA}H_A^{2} \tilde{E}_A^2
=1, ~~~{\rm or}~~~\tilde{E}_A={\rm const}
\,,
\label{eqn:Rii4}
\end{eqnarray}
Eqs. (\ref{eqn:Rii3}) and (\ref{traceless3})
are satisfied.

Since 
\begin{eqnarray}
M_{AA}=\frac{(q_A+1)(D-q_A-3)}{D-2}+\frac{1}{2}a_A^2
\equiv\frac{\Delta_A}{D-2}
\,,
\end{eqnarray}
from Eq. (\ref{eqn:Rii4}), we find 
 $\tilde{E}_A$ as
\begin{eqnarray}
\tilde{E}_A=\sqrt{\frac{2(D-2)}{\Delta_A}}\frac{1}{H_A}
\label{q-H2}, ~~~({\rm or}~~~\tilde{E}_A={\rm const}) 
\,.
\end{eqnarray}
If we impose that a spacetime is  asymptotically flat (i.e., $H_A
\rightarrow 1$ as $r\rightarrow \infty$) and the potential $E_A$
vanishes at infinity,
we find that
\begin{eqnarray}
E_A=-\sqrt{\frac{2(D-2)}{\Delta_A}}\left(1-\frac{1}{H_A}\right)
\label{q-H}, ~~~({\rm or}~~~E_A=0)
\,.
\label{eqn:potential}
\end{eqnarray}

Inserting this relation into Eq. (\ref{phi_A}), 
we obtain the equation
for $H_A$ as
\begin{eqnarray}
\partial^2 H_A=0
\,,
\label{harmonic_H}
\end{eqnarray}
which means that $H_A$ is a harmonic function on
$\{x^i\}\in\mathbb{E}^{d-1}$. From the relation (\ref{q-H}) with Eqs.
(\ref{eqn:Ruv}), (\ref{eqn:Raa}) and (\ref{eqn:phi}), we then obtain 
the solutions for metric functions in terms of the harmonic functions $H_A$:
\begin{eqnarray}
&&\xi=-\sum_A\frac{D-q_A-3}{\Delta_A}
\ln H_A,\quad\eta=\sum_A\frac{q_A+1}
{\Delta_A}\ln H_A,\nonumber\\
&&\zeta_\alpha=-\sum_A\frac{\delta_{\alpha A}}{\Delta_A}
\ln H_A,\quad\varphi=(D-2)\sum_A\frac{\epsilon_A a_A}{\Delta_A}\ln H_A
\,.
\label{sol:xietazetaphi}
\end{eqnarray}

We have two remaining equations (\ref{eqn:Rui}) and (\ref{eqn:maxwell})
 for ${\cal A}_i$ (${\cal F}_{ij}$) 
and one Poisson equation (\ref{eqn:Ruu}) for $f$ .
In order to solve the former two equations, we classify
 $n_A$-form field into  the following three cases:\\[1em]
(1) \underline{Charged Branes}\\[.5em]
We expect that each brane $A$ has a charge 
${\cal Q}_H^{(A)}$ (either electric
or magnetic type), and then 
$E_A$ becomes non-trivial, i.e., 
$H_A\neq 1$.
In this case,  if we set 
\begin{eqnarray}
B_i^A=-\tilde{E}_A {\cal A}_i=-{\sqrt{2(D-2)\over \Delta_A}}{{\cal A}_i\over H_A}
\,,
\end{eqnarray}
we have
\begin{eqnarray}
{\cal F}_{ij}^{(A)}={\sqrt{2(D-2)\over \Delta_A}}{\cal F}_{ij}
\,.
\label{FFA_relation0}
\end{eqnarray}
Inserting Eqs. (\ref{sol:xietazetaphi}),
we can show that two equations  (\ref{eqn:Rui}) and (\ref{eqn:maxwell})
are reduced to 
the following one Laplace equation:
\begin{eqnarray}
\partial^j {\cal F}_{ij} =0
\,.
\label{eq:A1}
\end{eqnarray}
$B_i^A$ describes a magnetic-type field produced by 
a current appearing through  rotation of a charged brane.

It turns out that the condition (\ref{FFA_relation0})
plays a key role for the system to keep
supersymmetry (see Appendix A).

~\\[1em]
(2) \underline{Neutral Branes with Currents}\\[.5em]
If $H_A=1$ (i.e., $E_A=0$),
 which is a trivial solution of Eq. (\ref{harmonic_H}),
we find that there is no electric type field, and then 
zero charge on the brane. This brane does not make any contribution to
$\xi,\eta,\zeta_\alpha$ and $\varphi$.
Although  
the electric type field vanishes,
 $B_i^A$ can still exist because a magnetic type field
is produced by a ``current", which is not charged.
This current may appear in  a system consisting of
the same numbers of branes and anti-branes, which move with different 
velocities.
This situation is similar to the conventional electric current
in a metal. The negative charges of electrons balance with
the positive ones of protons in the metal. The net charge vanishes,
but a current is produced by the motion of electrons.

In this case ($H_A=1$), 
the equations for ${\cal A}_i$ and $B_i^A$ become independent
as
\begin{eqnarray}
\partial^j {\cal F}_{ij} =0~~~~,~~~
\partial^j {\cal F}_{ij}^{(A)} =0
\,.
\label{harmonic:AB}
\end{eqnarray}
Since these two equations are exactly the same,
we can adopt the same solution with different amplitudes, i.e.,
\begin{eqnarray}
B^A_i=-\lambda_A\sqrt{\frac{2(D-2)}{\Delta_A} }{\cal A}_{i}\,,~~
{\cal F}_{ij}^{(A)}=\lambda_A{\sqrt{2(D-2)\over \Delta_A}}{\cal F}_{ij}
\,,
\label{FFA_relation1}
\end{eqnarray}
where
$\lambda_A$ is an arbitrary constant, which corresponds
to the strength of a current, i.e., numbers 
 of branes and anti-branes
and  its relative velocity.
This relation not only makes 
the equation for $f$ simple (see below)
but also keeps supersymmetry of the system 
  (see Appendix A) .

~\\
(3) \underline{Charged Branes with Currents}\\[.5em]
If numbers  of branes and anti-branes  are different,
such a system has a net charge.
Then the magnetic field may be divided into two parts:
\begin{eqnarray}
B^A_i=-\sqrt{\frac{2(D-2)}{\Delta_A}}{{\cal A}_{i}\over H_A}
+B_i^{A({\cal N})}
\,,
\end{eqnarray}
where the first term is produced by a current appearing
through the motion of  a 
net charge, while
$B_i^{A({\cal N})}$ is
produced by a current even in the case of
zero net charge (just as in Case (2)).
From Cases (1) and (2), we expect that ${\cal A}_{i}$
and $B_i^{A({\cal N})}$ are arbitrary harmonic functions
(just as Eqs. (\ref{harmonic:AB})).
However, since 
we have two equations (\ref{eqn:Rui}) and (\ref{eqn:maxwell})
for ${\cal A}_i$, it is not trivial whether our expectation
is the case. 
In fact, 
we have to impose
the following condition
\begin{eqnarray}
\partial_{[i}B^A_{j]}
\cdot \partial^j\tilde{E}_A=0\
\,,
\label{condition3}
\end{eqnarray}
in order for ${\cal A}_{i}$
and $B_i^{A}$
to be a solution.

If we can impose the relation of 
$B^{A({\cal N})}_i\propto {\cal A}_{i}$ 
just as that for $B^{A}_i$ in Case (2),
the equation for $f$ becomes simple,
but this relation is not consistent with
the condition (\ref{condition3}).
As a result, the equation for $f$ becomes 
very complicated,
although we can solve it in principle
because it is the Poisson equation
in a flat space.
We also find that this condition breaks the supersymmetry
of the system. Therefore, in what follows, 
we will discuss only Cases of (1) and (2).

Finally we discuss the last equation (\ref{eqn:Ruu})  for $f$.
Here we assume we have $N_{A'}$ charged branes
(Case (1))
and $N_{A''}$ neutral branes with currents (Case (2)).
$N_A=N_{A'}+N_{A''}$ is the total number of branes.
Then, 
as for the metric $f$, 
we find
\begin{eqnarray}
\partial^2 f={\beta\over 2} \prod_{A}H_{A}^{-\frac{2(D-2)}{\Delta_{A} }}
\left(\partial_{[j}{\cal A}_{i]}\right)^2
\,,
\label{Poisson:f}
\end{eqnarray}
where
\begin{eqnarray}
\beta=\left[1-(D-2)\left(
\sum_{A'} {1
\over \Delta_{A'} }
+\sum_{A''} {\lambda_{A''}^2
\over \Delta_{A''} }\right)
\right]
\,,
\label{def:beta}
\end{eqnarray}
is just a constant.
$A'$ describes charged branes which provide
non-trivial potentials $E_A$ (\ref{q-H2}),
while
$A''$ gives  a 
contribution from the neutral branes with currents.

If the following condition is satisfied:
\begin{eqnarray}
(D-2)\left(
\sum_{A'} {1
\over \Delta_{A'} }
+\sum_{A''} {\lambda_{A''}^2
\over \Delta_{A''} }\right) =1
\,,
\end{eqnarray}
$\beta$ vanishes, and then 
$f$ is given by an arbitrary harmonic function
on $\mathbb{E}^{d-1}$. 
In general, however, since $\lambda_{A''}$ is free, 
$\beta$ can have any 
sign (either positive, zero or negative).
Thus, we have to solve the Poisson equation (\ref{Poisson:f}).

The solution obtained in this section
is summarized as follows: 
\begin{eqnarray}
&&ds^2=\prod_A H_A^{2\frac{q_A+1}{\Delta_A}}
\left[2\prod_B H_B^{-2\frac{D-2}{\Delta_B}}
du\left(dv+fdu+{{\cal A}\over \sqrt{2}}\right)+\sum_{\alpha=2}^p\prod_B 
H_B^{-2\frac{\gamma_{\alpha B}}{\Delta_B}}dy_\alpha^2
+\sum_{i=1}^{d-1}dx_i^2\right],\nonumber\\&&
\varphi=(D-2)\sum_A \frac{\epsilon_A a_A}{\Delta_A} \ln H_A
\,,
\label{eqn:solution}
\end{eqnarray}
where 
\begin{eqnarray}
\Delta_A&=&(q_A+1)(D-q_A-3)
+\frac{D-2}{2}a_A^2,\nonumber\\ 
\gamma_{\alpha A}&=&\delta_{\alpha A}+q_A+1~~=~~\left\{\begin{array}{ll}
D-2 & ~~~\alpha=\alpha_2,\cdots, \alpha_{q_A} \\
0 & ~~~\mbox{otherwise}\end{array}\right. 
\,.
\end{eqnarray}
$H_A$  for each
$q_A$-brane and ${\cal A}={\cal A}_i dx^i$ are arbitrary harmonic
functions, while the vector potential
$B_i^A$ can be chosen either as $B_i^A\propto {\cal A}_i/H_A$
(when $H_A\neq 1$),
or an arbitrary harmonic
function (when $H_A=1$). 
The ``wave" metric
$f$  usually satisfies the Poisson equation (\ref{Poisson:f})
with some source term originated by the ``rotation"-induced
 metric ${\cal A}_i$,
although
it can be also an arbitrary harmonic function for some specific 
configuration of branes ($\beta=0$).

It is worth noticing that we have independent Laplace equations
for $H_A$ and ${\cal A}_i$ (and $f$ when $\beta=0$).
This makes the construction of solutions very easy.
The superposition of any solutions also provides us an exact solution.
Hence we can construct an infinite number of solutions.
We can also show that a part of supersymmetry is preserved
in Cases (1) and (2) 
if ${\cal F}_{ij}$ is self-dual (see Appendix A).

\section{``Black" Brane Solutions in M-theory and 
in type IIB superstring theory} 

In what follows, we shall show how to construct
the exact stationary 
solutions in M-theory (or in type IIB superstring
theory), and present concrete examples.
We do not know very much about M-theory or 
a superstring theory except for the 
fact that, at the low energy scale, it is 
described by  eleven-dimensional or ten-dimensional
supergravity. 
Then in this section, we discuss ``black" brane solutions in the
eleven-dimensional and ten-dimensional supergravity. 
Here we use the phrase ``black" brane 
for a stationary and asymptotically flat 
spacetime solution with intersecting branes,
although it may contain a naked singularity
or it will be a BPS solution for appropriate values of
parameters.

\subsection{M$2$ and M$5$-brane solutions in M-theory }

In 11-dimensional supergravity, we have a 4-form field
($n_A=4$) and 
no dilaton $\varphi$ ($a_A=0$).
Setting $D=11$ and $a_A=0$, we have   
\begin{eqnarray}
\Delta_A=(q_A+1)(8-q_A)
\,. 
\end{eqnarray}
The form field produced by an ``electric" charge is
related to the M2-brane, i.e., $q_A=n_A-2=2$. This gives
$\Delta_A=18$.
The ``black" brane solution in this case
is  written as
\begin{eqnarray}
ds_{11}^2&=&H_2^{1/3}\left[2H_2^{-1}du
\left(dv+fdu+{{\cal A}\over
\sqrt{2}}\right)
+H_2^{-1}dy_6^2+\sum _{i=1}^8dx_i^2\right],\nonumber\\
F_4&=&d(1/H_2)\wedge du\wedge dv\wedge dy_6 +{1\over \sqrt{2}}
dB_2\wedge du\wedge dy_6
\,,
\end{eqnarray}
where $H_2$ is a harmonic function on $\mathbb{E}^8$. 

Similarly,  the  field with a ``magnetic" charge 
is  related to the M5-brane because 
${}^*{q}_A={}^*{n}_A-2=D-n_A-2=5$. This also gives 
$\Delta _A=18$. The solution is described by
\begin{eqnarray}
ds_{11}^2&=&H_5^{2/3}\left[2H_5^{-1}du
\left(dv+fdu+{{\cal A}\over
\sqrt{2}}\right)
+H_5^{-1}\sum _{\alpha=2}^5dy_\alpha^2
+\sum _{i=1}^5dx_i^2\right],\nonumber\\
\ast F_4&=&d(1/H_5)\wedge du\wedge dv\wedge dy_2
\wedge dy_3\wedge dy_4\wedge dy_5
+{1\over \sqrt{2}}
dB_5\wedge du \wedge dy_2
\wedge dy_3\wedge dy_4\wedge dy_5
\,,
\end{eqnarray}
where $H_5$ is a harmonic function on $\mathbb{E}^5$. 
In both cases,
${\cal A}_i$ is also a vector harmonic function,
while
$f$ is given by the Poisson equation (\ref{Poisson:f})
with $\beta=1/2$
because of Eq. (\ref{def:beta})
(see the exact solution
in \S IV).

These two branes (M2 and M5) 
can  intersect if and only if 
\begin{eqnarray}
M2\cap M2\rightarrow \bar{q}_{22}=0,\quad
M2\cap M5\rightarrow \bar{q}_{25}=1,\quad
M5\cap M5\rightarrow \bar{q}_{55}=3
\,.
\end{eqnarray}
The crossing rule leads that there exists a
four-dimensional (4D) ``black" object with four independent branes 
(or three M5 branes and one wave), 
or a five-dimensional (5D) ``black" object with three independent M2 branes
(or two branes and one wave)
(see Table I). 
The 4 D ``black" object 
with  M2$\perp$M2$\perp$M5$\perp$M5 branes and 
the 5D object with M2$\perp$M2$\perp$M2  branes
have no traveling wave.
While, the 4D ``black" object with
M5$\perp$M5$\perp$M5$\perp$W  branes
and the 5D object with
M2$\perp$M5$\perp$W  branes describe stationary spacetimes with a 
traveling  wave. 
We shall discuss the details of  the 5D ``black" object with
M2$\perp$M5$\perp$W  branes in the next section.

\begin{table}[h]
\begin{center}
\begin{tabular}{|c|c|c|c|c|c|c|}
\multicolumn{7}{c}{$d=4$}\\ \hline
  $y_1$&$y_2$&$y_3$&$y_4$&$y_5$&$y_6$&$y_7$\\
  \hline
   M$2$ &  M$2$    &             &  &   &   &   \\
     &   & M$2$ &  M$2$      &             &   &  \\
   M$5$ &   & M$5$ &  & M$5$ &    M$5$      &    M$5$      \\
    & M$5$ &   & M$5$ & M$5$ &    M$5$      &    M$5$     \\
   \hline
\end{tabular}
\hskip .5cm
\begin{tabular}{|c|c|c|c|c|c|c|}
\multicolumn{7}{c}{$d=4$}\\   \hline
 $y_1$&$y_2$&$y_3$&$y_4$&$y_5$&$y_6$&$y_7$\\
  \hline
   M$5$ &             &             & M$5$ & M$5$ & M$5$ & M$5$ \\
   M$5$ & M$5$ & M$5$ &             &             & M$5$ & M$5$ \\
   M$5$ & M$5$ & M$5$ & M$5$ & M$5$ &             &             \\
    W &    &   &    &    &             &            \\
 \hline
\end{tabular}
\hskip .5cm
\begin{tabular}{|c|c|c|c|c|c|}
\multicolumn{6}{c}{$d=5$}\\ \hline
  $y_1$&$y_2$&$y_3$&$y_4$&$y_5$&$y_6$\\
  \hline
   M$2$ & M$2$   &    &     &      &  \\
     &   & M$2$  & M$2$   &   &     \\
     &    &   &   & M$2$  &M$2$  \\
   \hline
\end{tabular}
\hskip .5cm
\begin{tabular}{|c|c|c|c|c|c|}
\multicolumn{6}{c}{$d=5$}\\ \hline
  $y_1$&$y_2$&$y_3$&$y_4$&$y_5$&$y_6$\\
  \hline
   M$2$ &       &    &     &      & M$2$ \\
   M$5$ & M$5$ & M$5$ & M$5$ & M$5$ &     \\
     W &    &   &    &    &             \\
\hline
\end{tabular}
  \caption{Some examples of intersecting branes for $d=
4$ and 5. M2, M5 and W denote the location
where the M2 brane, the M5 brane and a wave exist, respectively.}
\label{intersection1}
\end{center}
\end{table}

\subsection{D1 and D5-brane solutions in type IIB superstring theory}
In the case of  ${\cal N}=2, 
D=10$ type IIB supergravity theory,
the action in the
 Einstein frame
is given by
\begin{eqnarray}
S=\frac{1}{2\kappa^2}\int
d^{10}X
\sqrt{-g}\left[{\cal R}
-\frac{1}{2}(\nabla\varphi)^2
-\frac{1}{2\cdot {n_A}!}e^{\frac{5-{n_A}}{2}\varphi}
F_{\bf n_A}^2\right]
\,.
\end{eqnarray}  
The coupling constant ${a_A}$ in the previous action (\ref{action0}) is
given by 
$a_A=(5-n_A)/2$.
The  three-form field with an ``electric" charge is
related to the D1-brane, i.e., $q_A=n_A-2=1$. 
Then we find
 $\Delta_A=(q_A+1)(7-q_A)
+4a_A^2=16$, which 
 does not 
depend on the type of branes ($A$ or ${n_A}$). 
This gives the same value of $(D-2)/\Delta_A=1/2$
as that in the case of eleven-dimensional supergravity.
Then we find solutions in type IIB supergravity similar to 
those in eleven-dimensional supergravity.
In fact a black brane solution in this case
is  written as
\begin{eqnarray}
&&ds_{10}^2=H_1^{1/4}\left[2H_1^{-1}du\left(dv+fdu+{{\cal A}\over \sqrt{2}}
\right)
+\sum _{i=1}^{8}dx_i^2\right]\nonumber\\
\,,
\end{eqnarray}
where $H_1$ is a harmonic function on $\mathbb{E}^8$.
The form field with a ``magnetic" charge  is  related to the D5-brane 
because ${}~*{q}_A={}~*{n}_A-2=D-n_A-2=5$. 
 This also gives 
$\Delta _A=16$. The solutions is described by
\begin{eqnarray}
ds_{10}^2=H_5^{-1/4}\left[
2du\left(dv+fdu+{{\cal A}\over \sqrt{2}}
\right)+H_5\sum_{i=1}^4dx_i^2
+\sum _{\alpha=2}^5dy_\alpha^2\right]
\,,
\end{eqnarray}
where $H_5$ is a harmonic function on $\mathbb{E}^4$. 
The solutions with two intersecting branes (D1 and D5)
are also given just as the previous subsection.

\subsection{The Compactification of a Black Brane}

The critical dimension for M-theory
(or a superstring ) is eleven (or ten).
We then have to compactify 
extra dimensions to obtain  an effective 
$d$-(four or 
five) dimensional world.

Rewriting the following part of the metric as
\begin{eqnarray}
2du\left(dv+fdu+{{\cal A}\over \sqrt{2}}
\right)&=&(1+f)\left[dy_1-\frac{1}{1+f}
\left(fdt-\frac{\cal A}{2}\right)\right]^2
-\frac{1}{1+f}\left(dt+\frac{\cal A}{2}\right)^2
\,,
\end{eqnarray}
we obtain our metric in $D$-dimensions as
\begin{eqnarray}
ds_D^2&=&\prod_A 
H_A^{2\frac{q_A+1}{\Delta_A}}
\left[-\prod_B H_B^{-2\frac{D-2}{\Delta_B}}\frac{1}{1+f}
\left(dt+\frac{\cal A}{2}\right)^2
+\sum_{i=1}^{d-1}dx_i^2\right]
\nonumber\\
&&+\prod_A H_A^{-2\frac{D-q_A-3}{\Delta_A}}(1+f)
\left[dy_1-\frac{1}{1+f}\left(fdt
-\frac{\cal A}{2}\right)\right]^2
+\sum_{\alpha=2}^p\prod_A H_A^{-2\frac{\delta_{\alpha A}}{\Delta_A}}
dy_\alpha^2 
\,.
\end{eqnarray}
Introducing the conformal factors $\Omega_1, \Omega_\alpha$ and $\Omega$ by
\begin{eqnarray}
&&
\Omega_1^2=(1+f)\prod_A H_A^{-2\frac{D-q_A-3}{\Delta_A}}
\nonumber \\
&&
\Omega_\alpha^2=\prod_A H_A^{-2\frac{\delta_{\alpha A}}{\Delta_A}}
~~~~~~(\alpha=2,\cdots, p)
\nonumber \\
&&
\Omega^2=\prod_{\alpha=1}^p \Omega_\alpha^2
=(1+f)\prod_A
H_A^{\frac{2}{\Delta_A}
\left[D-d-q_A(d-2)\right]}
\,,
\end{eqnarray}
we perform a conformal transformation of our metric as
\begin{eqnarray}
ds_D^2&=&\Omega^{-{2\over d-2}} d\bar{s}_d^2+\Omega_1^2 
\left[dy_1-\frac{1}{1+f}\left(fdt
-\frac{\cal A}{2}\right)\right]^2
+\sum_{\alpha=2}^p\Omega_\alpha^2 dy_\alpha^2
\,.
\end{eqnarray}
With this conformal transformation,
we obtain the Einstein gravity in $d$-dimensions, which 
 metric is given by 
\begin{eqnarray}
d\bar{s}_d^2&\equiv &\bar{g}_{\bar{\mu}\bar{\nu}}
dx^{\bar{\mu}} dx^{\bar{\nu}}
\nonumber \\
&=&
-\Xi^{d-3}\left(dt+\frac{\cal A}{2}\right)^2
+\Xi^{-1}\sum_{i=1}^{d-1}dx_i^2
\label{d_metric},\nonumber\\
\Xi&\equiv &(1+f)^{-1/(d-2)}
\prod_A H_A^{-\frac{2(D-2)}{(d-2)\Delta_A}}
\,,
\label{eqn:blackholes}
\end{eqnarray}
where $\bar{\mu}, \bar{\nu}, \cdots$
are coordinate indices for 
$d$-dimensional spacetime.
If the compactified space is sufficiently small,
we find the effective $d$-dimensional world with the metric
(\ref{d_metric}).

If this spacetime is asymptotically flat,
which we impose,
it may describe a ``black" object in
$d$-dimensions.
From the asymptotic form of the metric, we 
can define the ADM mass $M_{\rm ADM}$
as
\begin{eqnarray}
\bar{g}_{00}\sim -1+{16\pi G_d\over (d-2)\omega_{d-2}}
{M_{\rm ADM}\over r^{d-3}}
\,,
\label{ADM_mass}
\end{eqnarray}
where
\begin{eqnarray}
\omega_{d-2}\equiv
{2\pi^{d-1\over2}\over \Gamma\left({d-1\over2}\right)}, 
~~~{\rm and}~~~
r^2\equiv \sum_{i=1}^{d-1} x_i^2 
\,.
\end{eqnarray}
Assuming 
\begin{eqnarray}
H_A&\rightarrow&1+{{\cal Q}_H^{(A)}\over r^{d-3}}\nonumber \\
f&\rightarrow&{{\cal Q}_0\over r^{d-3}}
\,,
\end{eqnarray}
we obtain
\begin{eqnarray}
M_{\rm ADM}={(d-3)\pi^{d-3\over2}\over 8 G_d \Gamma\left({d-1\over2}\right)}
\left[{\cal Q}_0
+\sum_A{2(D-2)\over \Delta_A}{\cal Q}_H^{(A)}\right]
\,.
\label{def:ADM_mass}
\end{eqnarray}

For the case of eleven-dimensional M theory (and 
ten-dimensional type IIB string theory), 
we find
\begin{eqnarray}
\Xi&=&\left[(1+f)\prod_A{H_A}\right]^{-1/(d-2)}
\nonumber \\
M_{\rm ADM}&=&{(d-3)\pi^{d-3\over2}\over 8 G_d \Gamma\left({d-1\over2}\right)}
\left[{\cal Q}_0
+\sum_{A'}{\cal Q}_H^{(A')}\right]
\,,
\label{def:ADM_mass}
\end{eqnarray}
where $A'$ denotes charged branes.

Once we find solutions described by
the above set of equations, we have to 
 study a spacetime structure.
In particular, the horizon and the singularity of a spacetime
are important geometrical objects.
We then have to evaluate the curvature invariant
of the metric (\ref{eqn:blackholes}).
We calculate the Kretschmann invariant, which is given by
\begin{eqnarray}
\bar{{\cal R}}_{\bar{\mu}\bar{\nu}\bar{\rho}\bar{\sigma}}
\bar{{\cal R}}^{\bar{\mu}\bar{\nu}\bar{\rho}\bar{\sigma}}
&=&
{1\over 128}\Xi^{2(d-1)}
\left[3{\cal F}_{ij}^4+5({\cal F}_{ik}{\cal F}^{~k}_{j})^2
\right]
-{1\over 16}\Xi^{d}
\left[
4(\partial_i {\cal F}_{kl})^2
+{6(d-2)} \partial^i X~ 
 \partial_i({\cal F}_{kl}^2)
\right.
\nonumber\\
&-&
\left.
{4d} 
\partial^i\partial^j X   {\cal F}_{ik}{\cal F}^{~k}_{j}
+{2(3d^2-18d+22)}  (\partial^i X  {\cal F}_{ij})^2
+{(4d-9)} (\partial X )^2 {\cal F}_{ij}^2
\right]
\nonumber\\
&+&
{1\over 8}\Xi^2
\left[
{8(d-2)(d-3)}(\partial_i\partial_j X )^2+
8(\partial^2 X )^2+8{(d-2)^2(d-3)}\partial^i X   \partial^j X~
 \partial_i\partial_j X
\right.
\nonumber\\
&-&
\left.
8(d-2)(d-3)\partial^2 X(\partial X)^2
+(d-2)(d-3)(2d^2-8d+7)(\partial X)^4
\right]
\,,
\end{eqnarray}
where 
$X\equiv\ln \Xi
$.

For $d=5$,
we have
\begin{eqnarray}
\bar{{\cal R}}_{\bar{\mu}\bar{\nu}\bar{\rho}\bar{\sigma}}
\bar{{\cal R}}^{\bar{\mu}\bar{\nu}\bar{\rho}\bar{\sigma}}
&=&
{\Xi^8\over 128}\left[3{\cal F}_{ij}^4+5({\cal F}_{ik}{\cal F}^{~k}_{j})^2
\right]
\nonumber\\
&-&
{\Xi^5\over 16}
\left[
4(\partial_i {\cal F}_{kl})^2
+{18}\partial^i X \partial_i({\cal F}_{kl}^2)
-{20}\partial^i\partial^j X {\cal F}_{ik}{\cal F}^{~k}_{j}
+{14} (\partial^i X {\cal F}_{ik})^2
+{11} (\partial X)^2 {\cal F}_{ij}^2
\right]
\nonumber\\
&+&{\Xi^2\over 4}\left[
24(\partial_i\partial_j X)^2+ 
4 (\partial^2 X)^2+72 \partial^i  X \partial^j   X
 \partial_i\partial_j X
-24(\partial  X)^2\partial^2 X
+
{51}(\partial X)^4
\right]
\,.
\end{eqnarray}

In what follows,
we present the exact solutions for $D=11$ and $d=5$.
For $D=10$, 
the construction of solutions is almost the same as that of $D=11$.

\section{``black" brane solutions with M2-M5 (D1-D5) branes : 
the case of $d=5$}

We consider solutions in five-dimensions.
There are two branes (M2 and M5).
Then $N_{A'}+N_{A''}=2$.
In the ten-dimensional type IIB case,
we find the exactly the same as 
what we show below,
when we replace M2 and M5 
with
D1 and D5
(the indices $A=2,5$ with the indices $A=1,5$).

The metric in five-dimensions
is written by
\begin{eqnarray}
d\bar{s}_5^2=-\Xi^2
\left(dt+\frac{\mathcal{A}}{2}
\right)^2+\Xi^{-1}ds_{\mathbb{E}^4}^2
\,,
\end{eqnarray}
where 
$\Xi=\left[H_2H_5(1+f)\right]^{-1/3}$.
The unknown functions $H_A (A=2,5)$,
 ${\cal A}_i$ and $f$ satisfy
the following equations:
\begin{eqnarray}
&&\partial^2 H_A=0 ~~~(A=2,5)
\label{eqn:H}\\
&&\partial_j {\cal F}^{ij}=0
\label{eqn:A}\\
&& \partial^2 f={\cal S}\equiv {\beta \over 8H_2H_5 }
{\cal F}^{ij}{\cal F}_{ij}
\,,
\label{eqn:f}
\end{eqnarray}
where
\begin{eqnarray}
&&{\cal F}_{ij}=\partial_i{\cal A}_j-\partial_j{\cal A}_i
\nonumber \\
&&\beta =1-{1\over 2}\left(N_{A'}+\sum_{A''}\lambda_{A''}^2\right)
\,,
\end{eqnarray}
which value is explicitly given in Table \ref{beta}.

\begin{table}[h]
\begin{center}
\begin{tabular}{|c||c|c|c||c|c|}
 \hline
&&&&&\\[-1em] 
 & ${\rm M}_{2}$ &${\rm M}_{5}$&type of source branes  
&$\beta$&$H_A$\\[.5em]
  \hline
  \hline
&&&&&\\[-1em] 
(1)& ${\cal C}$&${\cal C}$ &two charged branes &0&$H_2, H_5$ : harmonic functions\\[.5em]
  \hline
&&&&&\\[-1em] 
(2a)&
${\cal C}$&${\cal N}$
 &charged \& neutral  branes &${1\over 2}\left(1-\lambda_5^2\right)$&$H_2 
$ : a harmonic function, $H_5=1$ 
\\[.5em]
  \hline
&&&&&\\[-1em] 
(2b)&
${\cal N}$&${\cal C}$ &neutral \& charged branes &${1\over 2}\left(1-\lambda_2^2\right)$ 
&$H_2=1$, $H_5$: a harmonic function \\[.5em]
  \hline
&&&&&\\[-1em] 
(2c) &${\cal N}$&${\cal N}$&
two neutral  branes &$1-{1\over 2}\left(\lambda_2^2+\lambda_5^2\right)$
&$H_2=H_5=1$ \\[.5em]
  \hline
\end{tabular}
  \caption{
The type of source branes and the value of $\beta$. There are
two branes (M2 and M5).
${\cal C}$ and ${\cal N}$ denote
a charged brane and a neutral brane with a current. $\lambda_2$
and $\lambda_5$
 are arbitrary parameters, which correspond to  current strength.
 }
\label{beta}
\end{center}
\end{table}

In order to find the exact solutions, we assume that
the 4-dimensional $x$-space has two rotation symmetries which
 Killing vectors ($\xi_{(\phi)}^i$ and $\xi_{(\psi)}^i$) commute
each other.
In this case, Eq. (\ref{eqn:A}) is reduced to two
uncoupled equations for two scalar fields,
${\cal A}_\phi={\cal A}_i \xi_{(\phi)}^i$ and 
${\cal A}_\psi={\cal A}_i \xi_{(\psi)}^i$, as
\begin{eqnarray}
&&
\partial^2 {\cal A}_\phi
-\partial_i \ln \left(\xi_{(\phi)}\cdot\xi_{(\phi)}
\right)~\partial^i {\cal A}_\phi
=0\,,
\label{eqn:A_phi}\\
&&
\partial^2 {\cal A}_\psi
-\partial_i \ln \left(\xi_{(\psi)}\cdot\xi_{(\psi)}
\right)~\partial^i {\cal A}_\psi
=0
\,.
\label{eqn:A_psi}
\end{eqnarray}
Here we have assumed that 
the other components of ${\cal A}_i$ vanish. 

We now have the Laplace equations or similar
equations (the Poisson equation or Eqs. (\ref{eqn:A_phi}) and 
(\ref{eqn:A_psi})) 
for several scalar functions ($H_A$, ${\cal A}_\phi$, 
${\cal A}_\psi$, and $f$).
Each equation is linear and 
uncoupled except for the equation for $f$ with $\beta\neq 0$.
Hence it is very easy to find general solutions
because the Laplace-Beltrami operator is
defined on the flat Euclidian space.
Once we obtain a complete set of solutions
in an appropriate curvilinear coordinate system,
we can construct any solutions by superposing them.

Giving an explicit form of a solution, we 
obtain the properties of a ``black" object.
For example, 
assuming the asymptotic behaviors for $H_A$ and $f$
as
\begin{eqnarray}
H_A&\rightarrow&1+{{\cal Q}_H^{(A)}\over r^2}\nonumber \\
f&\rightarrow&{{\cal Q}_0\over r^2}
\,,
\end{eqnarray}
as 
\begin{eqnarray}
r\equiv\left({\sum_{i=1}^4 x_i^2}\right)^{1/2}\rightarrow \infty
\,,
\end{eqnarray}
we find
\begin{eqnarray}
M_{\rm ADM}={\pi\over 4G_5}\left({\cal Q}_0+{\cal Q}_H^{(2)}
+{\cal Q}_H^{(5)}
\right)
\,.
\end{eqnarray}

The entropy of a black hole, if it exists,
is defined by
\begin{eqnarray}
S&=&{A_h\over 4G_5}
\,,
\end{eqnarray}
where $ A_h$ is the area of horizon. 

In what follows, adopting the hyperspherical coordinates
as a curvilinear coordinate system, we show explicitly
how to construct the exact solutions.

\subsection{hyperspherical coordinates}
We adopt the hyperspherical coordinates:
\begin{eqnarray}
x_1+ix_2=r\cos\theta e^{i\phi},\quad 
x_3+ix_4=-r\sin\theta e^{i\psi}
\,,
\end{eqnarray}
where $0\leq \phi, \psi <2\pi$ and $0\leq \theta \leq \pi/2$.
The line element of 4D flat space is
\begin{eqnarray}
ds_{\mathbb{E}^4}^2=dr^2+r^2\left(d\theta^2+ \cos^2\theta d\phi^2
+ \sin^2\theta d\psi^2\right)
\,.
\end{eqnarray}
The symmetric axis is described by $\theta=0$ and $\pi/2$, and
the infinity corresponds to $r= \infty$.

Eq. (\ref{eqn:H}) in this coordinate system is  
\begin{eqnarray}
{1\over r}\partial_r\left(r^3
\partial_r H_A \right)+
{1\over \sin\theta\cos\theta }
\partial_\theta\left(\sin\theta\cos\theta
\partial_\theta H_A \right)=0
\,. 
\label{eqn:H_sphere}
\end{eqnarray}
Setting $H_A=h_A(r)j_A(\theta)$,
we  separate the variables and obtain two ordinary differential equations:
\begin{eqnarray}
&&{1\over r}{d\over dr}\left(r^3
{dh_A\over dr} \right)-Mh_A=0\,,
\label{eqn:H_sphere1}\\
&&{1\over \sin\theta\cos\theta }
{d\over d\theta}\left(\sin\theta\cos\theta
{d j_A\over d\theta} \right)+Mj_A=0
\,,
\label{eqn:H_sphere2}
\end{eqnarray}
where $M$ is a separation constant.
Eq. (\ref{eqn:H_sphere2}) with $\mu=\cos 2\theta $ 
is just the Legendre equation
as
\begin{eqnarray}
{d\over d\mu}\left((1-\mu^2){dj_A\over d\mu}\right)+
{M\over 4}j_A=0
\,.
\label{eqn:H_Legendre}
\end{eqnarray}
From regularity conditions on the symmetric axis ($\theta=0,\pi/2$),
we obtain $j_A=P_\ell(\cos 2\theta )$ by setting 
$M=4\ell (\ell+1)$ ($\ell=0,1,2, \cdots$).
Eq. (\ref{eqn:H_sphere1})
is easily solved as $h_A=r^{2\ell}$ or $r^{-2(\ell+1)}$.
The general solution for $H_A$ is then
\begin{eqnarray}
H_A=\sum_{\ell=0}^\infty
\left[g_\ell^{(A)} r^{2\ell}+h_\ell^{(A)} r^{-2(\ell+1)}\right] 
P_\ell(\cos 2\theta)
\,,
\label{sol:H_sphere}
\end{eqnarray}
where $g_\ell^{(A)}$ and $h_\ell^{(A)}$ are arbitrary constants.

>From the asymptotically flatness condition, the solution is given by
\begin{eqnarray}
H_A=1+\sum_{\ell=0}^\infty h_\ell^{(A)}
 r^{-2(\ell+1)}P_\ell (\cos 2\theta )
\,.
\end{eqnarray}

The spherically symmetric solution ($\ell=0$) is given by
\begin{eqnarray}
H_A=1+{{\cal Q}_H^{(A)}\over r^2}
\,,
\label{HA_spherical}
\end{eqnarray}
where ${\cal Q}_H^{(A)}$ is a constant, which corresponds to 
a conserved charge.

Next, we discuss Eqs. (\ref{eqn:A_phi}) and (\ref{eqn:A_psi}), which are 
written as
\begin{eqnarray}
&&r \partial_r\left(r
\partial_r {\cal A}_\phi \right)+
\cot \theta 
\partial_\theta\left(\tan \theta
\partial_\theta {\cal A}_\phi \right)=0\,,
\label{eqn:Aphi_sphere}\\
&&r \partial_r\left(r
\partial_r {\cal A}_\psi \right)+
\tan \theta 
\partial_\theta\left(\cot \theta
\partial_\theta {\cal A}_\psi \right)=0
\,.
\label{eqn:Apsi_sphere}
\end{eqnarray}
Setting ${\cal A}_\phi=a_\phi(r) b_\phi (\theta)$ and 
${\cal A}_\psi=a_\psi(r) b_\psi (\theta)$, we have the following 
ordinary differential equations:
\begin{eqnarray}
&&r {d \over dr}\left(r
{d\over dr} a_\phi \right)
-Ka_\phi=0
\label{eqn:Aphi_sphere1}\\
&&{d^2b_\phi\over d\mu^2}-{1\over 1-\mu}{d b_\phi\over d\mu}
+{K\over 4(1-\mu^2)}b_\phi=0\,,
\label{eqn:Aphi_sphere2}\\
&&r {d \over dr}\left(r
{d\over dr} a_\psi \right)
-La_\psi=0
\label{eqn:Apsi_sphere1}\\
&&{d^2b_\psi\over d\mu^2}+{1\over 1+\mu}{d b_\psi\over d\mu}
+{L\over 4(1-\mu^2)}b_\psi=0
\,,
\label{eqn:Apsi_sphere2}
\end{eqnarray}
where $\mu=\cos 2\theta$, and $K$ and $L$ are
separation constants.
The solutions of Eqs. (\ref{eqn:Aphi_sphere2}) and (\ref{eqn:Apsi_sphere2}) 
are described by Gauss's hypergeometric functions as
$b_\phi(\mu)=F(-\sqrt{K}/2,\sqrt{K}/2, 1, (1-\mu)/2)$ and 
$b_\psi(\mu)=F(-\sqrt{L}/2,\sqrt{L}/2, 1, (1+\mu)/2)$.
The Gauss's hyper geometrical function $F(\alpha,\beta,\gamma,z)$ is 
defined by 
\begin{eqnarray}
F(\alpha,\beta,\gamma,z)=\frac{\Gamma(\gamma)}
{\Gamma(\alpha)\Gamma(\beta)} \sum_{n=0}^\infty
\frac{\Gamma(\alpha+n)\Gamma(\beta+n)}{\Gamma(\gamma+n)}
\frac{z^n}{n!}
\,.
\end{eqnarray}
From regularity conditions, we have to impose that
$K=4m^2$ and $L=4n^2$, where $m, n =1,2, \cdots$.
We then have the angular solutions as 
$b_\phi=F(-m, m, 1, \sin^2\theta)$ and 
$b_\psi=F(-n, n, 1, \cos^2\theta)$.
The explicit forms of this hypergeometric function with $m,n=1,2$ 
are given by
\begin{eqnarray}
F(-1,1,1,z)&=&1-z
\nonumber \\
F(-2,2,1,z)&=&1-4z+3z^2
\,.
\end{eqnarray}

The equations for $a_\phi$ and $a_\psi$ are easily solved,
i.e.,  $a_\phi=r^{2m}, r^{-2m}$ and 
$a_\psi=r^{2n}, r^{-2n}$.
We then obtain a general solution for ${\cal A}_i$ as
\begin{eqnarray}
&&{\cal A}_\phi=\sum_{m=1}^\infty
\left [a^{(\phi)}_m r^{2m}+b^{(\phi)}_m r^{-2m}\right] F(-m,m,1,\sin^2\theta)  
\label{sol:Aphi_sphere}\\
&&{\cal A}_\psi=\sum_{n=1}^\infty
\left[a^{(\psi)}_n r^{2n}+b^{(\psi)}_n r^{-2n}\right] F(-n,n,1,\cos^2\theta) 
\,,
\label{sol:Apsi_sphere}
\end{eqnarray}
where $a^{(\phi)}_m$, $b^{(\phi)}_m$,
$a^{(\psi)}_n$ and $b^{(\psi)}_n$ are arbitrary constants.

Assuming  asymptotically flatness, the solution for ${\cal A}_i$ is given by 
\begin{eqnarray}
&&{\cal A}_\phi=\sum_{m=1}^\infty {b^{(\phi)}_m\over r^{2m}}
F(-m,m,1,\sin^2\theta),\\
&&{\cal A}_\psi=\sum_{n=1}^\infty {b^{(\psi)}_n\over r^{2n}}
F(-n,n,1,\cos^2\theta)
\,.
\end{eqnarray}

If we take the first two terms in the general solution,
we obtain a simple solution as
\begin{eqnarray}
{\cal A}_\phi
&=&{\cos^2\theta\over r^2}\left[J_1^{(\phi)}+
{J_2^{(\phi)}\over r^2}(1-3\sin^2\theta)\right]
\label{sol:A_phi}
 \\
{\cal A}_\psi
&=&{\sin^2\theta\over r^2}\left[J_1^{(\psi)}+
{J_2^{(\psi)}\over r^2}(1-3\cos^2\theta)\right]
\,,
\label{sol:A_psi}
\end{eqnarray}
where $J_1^{(\phi)},J_1^{(\psi)},J_2^{(\phi)}$ and $J_2^{(\psi)}$
 are constants. The first two constants describe angular momenta
of a ``black" object.
As we show in Appendix A, if ${\cal F}_{ij}$ is self-dual,
the spacetime is supersymmetric. This condition implies $J_1^{(\phi)}
=-J_1^{(\psi)}$ and $J_2^{(\phi)}=J_2^{(\psi)}$.

Finally we discuss  Eq. (\ref{eqn:f}):
\begin{eqnarray}
&&\frac{1}{r}\partial_r(r^3\partial_rf)+\frac{1}{\sin\theta\cos\theta}
\partial_\theta(\sin\theta\cos\theta\partial_\theta f)
={\cal S}(r,\theta)\nonumber\\
&&\equiv {\beta \over 4 H_2H_5}\left[
{1\over \cos^2\theta}\left((\partial_r{\cal A}_\phi)^2+{1\over r^2}
(\partial_\theta{\cal A}_\phi)^2\right)
+{1\over \sin^2\theta}\left((\partial_r{\cal A}_\psi)^2+{1\over r^2}
(\partial_\theta{\cal A}_\psi)^2\right)
\right]
\,.
\end{eqnarray}

\subsection{BMPV type solutions: $\beta=0$}

If $\beta=0$, i.e., Case (1) (two charged brane)
or Case (2a-2c) (neutral branes) with
appropriately chosen current strength $\lambda_{A''}$,
 we find the Laplace equation for $f$,
which gives us a simple solution:
\begin{eqnarray}
f=\sum_{\ell=0}^\infty
{\cal Q}_\ell r^{-2(\ell+1)}
P_\ell(\cos 2\theta)
\,,
\label{sol:f_sphere}
\end{eqnarray}
where ${\cal Q}_\ell$'s are constants.

In this case, the solution with the lowest multipole moment is
given by
\begin{eqnarray}
&&H_A=1+{{\cal Q}_H^{(A)}\over r^2}~~~~(A=2,5)\,,
\nonumber \\
&&f={{\cal Q}_0 \over r^2}\,,
\nonumber \\
&&{\cal A}_\phi
={J_\phi\cos^2\theta\over r^2}\,,
\nonumber \\
&&{\cal A}_\psi
={J_\psi\sin^2\theta\over r^2}
\,.
\label{BMPV}
\end{eqnarray}
The mass and the entropy of this spacetime are 
\begin{eqnarray}
M_{\rm ADM}&=&\frac{\pi}{4G_5}({\cal Q}_0+{\cal Q}_H^{(2)}+{\cal Q}_H^{(5)}),\\
S&=&{A_h\over 4G_5}~=~
\frac{\pi^2}{3G_5} {\Lambda_+^2+\Lambda_+\Lambda_-
+\Lambda_-^2\over \Lambda_+^{3/2}+\Lambda_-^{3/2}}
\,,
\end{eqnarray}
where
\begin{eqnarray}
\Lambda_+&=&{\cal Q}_0{\cal Q}_H^{(2)}{\cal Q}_H^{(5)}-{J^2\over 8}
+{\Delta J^2\over 16}\,,\\
\Lambda_-&=&{\cal Q}_0{\cal Q}_H^{(2)}{\cal Q}_H^{(5)}-{J^2\over 8}
-{\Delta J^2\over 16}
\,.
\end{eqnarray}
$J^2$ and $\Delta J^2$ are defined by $J^2\equiv (J_\phi^2+J_\psi^2)/2$
 and $\Delta J^2\equiv J_\phi^2-J_\psi^2$, respectively.

Fixing $J^2$, if we maximize entropy $S$, we find the 
maximum entropy with 
\begin{eqnarray}
S=S_{\rm max}=
\frac{\pi^2}{2G_5} \sqrt{
{\cal Q}_0{\cal Q}_H^{(2)}{\cal Q}_H^{(5)}-{J^2\over 8}}
\,,
\end{eqnarray}
 if $\Delta J^2=0$, i.e., $J_\phi^2=J_\psi^2=J^2$.
Note that supersymmetry implies $J_\phi=-J_\psi=J$,
which corresponds to the BMPV solution \cite{BMPV,herdeiro}.
If $J_\phi\neq -J_\psi$, the above solution describes
a regular rotating non-BPS black hole spacetime in five dimensions.

\subsection{Brinkmann wave type solutions: $\beta\neq 0$}
When $\beta\neq 0$, 
since the source term is quadratic with respect to ${\cal A}_i$, 
it is not so easy to
find a general solution.
However, once we know the explicit form of the source term,
expanding $f(r,\theta)$ and the source term
${\cal S}(r,\theta)$ 
by the Legendre functions as 
\begin{eqnarray}
f(r, \theta)&=&\sum_{\ell=0}^\infty f_\ell(r)P_\ell(\cos 2\theta)\,,
\label{expand_f_Legendre}
\nonumber \\
{\cal S}(r,\theta)&=& \sum_{\ell=0}^\infty {\cal S}_\ell(r)
P_\ell(\cos 2\theta)
\,,
\label{expand_S_Legendre}
\end{eqnarray}
we find the ordinary differential equation for each moment $\ell$ as
\begin{eqnarray}
{1\over r}{d\over dr}\left(r^3{df_\ell(r)\over dr}\right) 
-4\ell(\ell+1)f_\ell(r)
={\cal S}_\ell(r)
\,.
\end{eqnarray}
If we can integrate this equation, we find an analytic solution.

Here we give one simple example, i.e.,
$H_2=H_5=1$ [Case (2c) in Table II] with Eqs. (\ref{sol:A_phi}) and 
(\ref{sol:A_psi}).
We find the solutions as
\begin{eqnarray}
f=f_0(r)+f_1(r)P_1(\cos 2\theta) +f_2(r) P_2(\cos 2\theta)
\,,
\end{eqnarray}
with
\begin{eqnarray}
f_0&=&{{\cal Q}_0\over r^2}+\beta\left({J_1^2 \over 12r^6}
+{ J_2^2\over 
20r^{10}}\right)
\nonumber \\
f_1&=&{{\cal Q}_1\over r^4}+\beta{J^{(\phi\psi)}_{12}
\over 40r^8}
\nonumber \\
f_2&=&{{\cal Q}_2\over r^6}+\beta {J_2^2 \over 14r^{10}}
\,,
\end{eqnarray}
where ${\cal Q}_0,{\cal Q}_1,{\cal Q}_2$ are 
integration constants and $2J_1^2=
(J_1^{(\phi)})^2+(J_1^{(\psi)})^2$, $2J_2^2=
(J_2^{(\phi)})^2+(J_2^{(\psi)})^2$, and $\Delta J^{(\phi\psi)}_{12}
=J_1^{(\phi)}J_2^{(\phi)}-J_1^{(\psi)}J_2^{(\psi)}$.

If we set $J_1^{(\phi)}=-J_1^{(\psi)}=J$ and $J_2^{(\phi)}=J_2^{(\psi)}=0$,
we find
\begin{eqnarray}
H_2&=&H_5~=~1
\nonumber \\
f&=&{{\cal Q}_0\over r^2}+\beta{J^2\over 12r^6}
\nonumber \\
{\cal A}_\phi&=&{J \over r^2}\cos^2\theta\,,
\nonumber \\
{\cal A}_\psi&=&-{J \over r^2}\sin^2\theta
\label{sol:Brinkmann}
\,.
\end{eqnarray}
We then recover
 the Brinkmann solution 
by setting $\beta=1$ (i.e., $B_i^A=0$) \cite{Brinkmann}.
If $\beta<0$, $(1+f)$ vanishes at $r=r_S (>0)$, which is a singularity.
For the case of $\beta>0$, this solution is similar to
the Brinkmann wave.

We can extend the above solution to Case (2a) in Table II:
 $H_2\neq 1$ and $H_5=1$ (or (2b):$H_2 =1$ and $H_5\neq 1$).
We find the following new solution.
Supposing that $H_2$  depends only on $r$ as
Eq. (\ref{HA_spherical})
and the lowest moment for ${\cal A}_i$,
we can obtain the exact solution:
\begin{eqnarray}
H_2&=&1+{{\cal Q}_H^{(2)}\over r^2}\\
f&=&{Q_0\over r^2}+{\beta J^2\over 4r^6(H_2-1)^3}
\left[H_2^2-1-2H_2\ln H_2
\right]\\
{\cal A}_\phi 
&=&{J_1^{(\phi)} \over r^2}\cos^2\theta\\
{\cal A}_\psi 
&=&{J_1^{(\psi)} \over r^2}\sin^2\theta
\,,
\end{eqnarray}
where $2J^2=\left[\left(J_1^{(\phi)}\right)^2+\left(J_1^{(\psi)}\right)^2\right]$.
The asymptotic behavior of this solution is
\begin{eqnarray}
f&\rightarrow&{{\cal Q}_0\over r^2}+\beta{J^2\over 12r^6}
~~~({\rm as}~~r\rightarrow \infty)
\nonumber \\
&\rightarrow&{\beta J^2\over 4{\cal Q}_H^{(2)}r^4}
~~~({\rm as}~~r\rightarrow 0)
\,.
\end{eqnarray}
The ADM mass is given by
\begin{eqnarray}
M_{\rm ADM}&=&\frac{\pi}{4G_5}({\cal Q}_0+{\cal Q}_H^{(2)})
\,.
\end{eqnarray}
Although $r=0$ is not a singularity,
it is not a horizon because it is 
a timelike hypersurface.
In fact, setting $J_1^{(\phi)}=-J_1^{(\psi)}=J$
(a supersymmetric spacetime),
we find the surface area of $r$=constant
as
\begin{eqnarray}
A(r)&=&4\pi r^3\int d\theta \cos\theta\sin\theta\left[H_2(1+f)-{J^2\over 8r^6}
\right]^{1/2}
\nonumber
\\
&\sim& \pi\int d\theta \cos\theta\sin\theta ~
|J|\left(
\beta-{1\over 2}
\right)^{1/2} ~~~({\rm as}~r\rightarrow 0)
\,.
\end{eqnarray}
This value  becomes imaginary because
$\beta<1/2$.
When $\beta=1/2$, i.e., $B_i^A=0$,
 the surface area (the entropy)
vanishes.

This solution is also similar to the Brinkmann wave solution
if $\beta>0$.
For $\beta=0$, we have already discussed in the previous subsection.
In the case of $\beta<0$,
$(1+f)$ vanishes at finite radius $r=r_S(>0)$,
and then there exists a naked singularity.

\section{Concluding Remarks}

In this paper, we have studied a stationary ``black" brane solution
in M/superstring theory.
Assuming a BPS type relation between the first-order derivatives
of metric functions, 
we have shown how to construct a stationary ``black" brane solution 
 with a traveling wave. 
We consider two types of intersecting brane: (1) charged branes and 
(2) neutral branes with a current.
The solutions are given by harmonic functions
$H_A$ and ${\cal A}_i$
plus a wave metric $f$ which satisfies the Poisson equation
for $\beta\neq 0$ (Cases (2a)-(2c) in Table II)
or the Laplace equation for $\beta=0$ (Case (1) and Cases
(2a)-(2c) with specific values
for $\lambda_A$ in Table II).
Since those differential equations are linear and independent 
except for  the Poisson equation for $f$,
we can easily construct general solutions by superposition
of harmonic functions.

Using the hyperspherical coordinate system,
we  present exact solutions in eleven-dimensional
M theory 
for the case with M2$\perp$M5  intersecting branes
and a traveling wave.
Compactifying these solutions
into five dimensions, we show that these solutions
 include  the BMPV black hole 
and the Brinkmann wave solution.
We have also found new solutions  which are similar to the Brinkmann wave.

We have proved that the solutions preserve the 1/8 supersymmetry 
if ${\cal F}_{ij}$ is self-dual.
All solutions found in the hyperspherical coordinates
preserve the 1/8 supersymmetry if the angular momenta
satisfy some relation (e.g., $J_\phi=-J_\psi$).

We also discuss 
non-spherical ``black" brane solutions (e.g., a ring topology and 
 an elliptical shape solution)
by use of hyperelliptical and hyperpolorical coordinates
in Appendix B.
Unfortunately, we could not find 
regular new solution, but in stead, solutions with a naked singularity.
In particular, even if we use  hyperpolorical coordinates,
we could not obtain a supersymmetric black ring.
We may have to extend our approach to find such a black ring solution.
We have two possibilities: One is the extension of our metric form, 
and the other is different configuration of intersecting branes.
Bena and Kraus \cite{Bena_Kraus} describe a black ring solution in M-theory.
If we rewrite the solution in our null coordinate system,
we find that we have to include another ``gravi-electromagnetic" field
in addition to ${\cal A}_j$.
While Elvang et al. \cite{EEMR05_1} 
use  three charged M2 branes  and three m5 monoploles
to describe a supersymmetric black ring.
Thus we may have to consider a different type of
brane configuration.
Using  the  
U-duality  \cite{lunin}, we may construct a supersymmetric black 
ring \cite{EEMR05_1} and other solutions
from our brane solutions. 

 The  charges of branes   of the 
 BMPV black hole  correspond  to the numbers of D-brane 
 tension.
While SO(4) rotational symmetries,
which describe angular momenta of the black hole,
corresponds to  endmorphisms in the graded algebra that rotate the 
 fermionic generators $G^i_m$  \cite{herdeiro}. 
By this correspondence (AdS/CFT correspondence), we can discuss
the properties of our solutions in the SCFT side.

Although we assume the BPS type relations for the metric,
we have to solve the elliptic type differential equations
if we want to find most general solutions, especially
non-BPS spacetimes. For this purpose, 
we need a completely different approach 
 such as a soliton technique to generate
new solutions \cite{Belinski_Zakharov1, Belinski_Zakharov2, Mishima_Iguchi}.

We have found that the BPS and non-BPS rotating 
asymptotically flat stringy black holes,  from which 
we may learn more about connections between microscopic and  macroscopic 
states of gravitating objects. 
In our framework, we consider a toroidally compactified string 
theory, but one may embed the BMPV type geometry in 
M-theory compactified on generic  Calabi-Yau spaces,
which would be more interesting.

\section*{Acknowledgments}
We would like to thank Gary Gibbons, Renata Kallosh,
Tatsuhiko Koike, Shuntaro Mizuno, 
Hidefumi Nomura, Masato Nozawa, Nobuyoshi Ohta, and 
Takashi Torii for useful discussions. 
KM also acknowledges Nobuyoshi Ohta 
for his especially helpful comments and MT expresses his heartfelt thanks 
to Hiroyuki Hyuga for encouragement.
This work was partially supported by the Grant-in-Aid for Scientific Research
Fund of the JSPS (No. 17540268) and another for the 
Japan-U.K. Research Cooperative Program,
and by the Waseda University Grants for Special Research Projects and 
 for The 21st Century
COE Program (Holistic Research and Education Center for Physics
Self-organization Systems) at Waseda University.

\appendix
\section{Supersymmetry in  Eleven-dimensional ``Black" Branes}

Here we discuss supersymmetry in the solution obtained in 
this paper.
The invariance for supersymmetry transformation
of a gravitino gives a criterion for  existence of 
unbroken supersymmetry. This condition is given by
 the Killing equation
for the Killing spinor $\epsilon$  \cite{CJS}, i.e.,
\begin{eqnarray}
\delta\psi _{\hat A}=\left[e^{~\mu}_{\hat A}\partial _\mu 
+\frac{1}{4}{w^{\hat B\hat C}}_{\hat A}\gamma _{\hat B\hat C}
+\frac{1}{288}({\gamma _{\hat A}}^{~\hat B\hat C\hat D\hat E}
-8\delta^{~\hat B}_{\hat A}\gamma^{\hat C\hat D\hat E})
F_{\hat B\hat C\hat D\hat E}\right]\epsilon=0
\,,
\label{eqn:Killingspinor}
\end{eqnarray}
where $\gamma_{\hat A\hat B}$'s are the antisymmetrized products of 
eleven-dimensional gamma matrices with unit strength on 
vielbein ${e^{\hat A}}_{~\mu}$, and a spin connection is given by
\begin{eqnarray}
{w^{\hat B\hat C}}_{\hat A}=e_{\hat A}^{~~\mu}\left(e^{\hat B\nu}\partial_{[\mu}
{e^{\hat C}}_{\nu]}-e^{\hat C\nu}\partial_{[\mu}{e^{\hat B}}_{\nu]}
-e^{\hat B\rho}e^{\hat C\sigma}e_{\hat D\mu}
\partial_{[\rho}{e^{\hat D}}_{\sigma]}\right)
\label{dif:spinconection}
\,.
\end{eqnarray}

Now we consider the M2$\perp$M5  ``black" brane solution 
related to the five-dimensional black hole, which 
we have discussed in section IV. 
The metric functions $\xi,\eta,\zeta_\alpha$ for
the space with M2$\perp$M5  `intersecting branes are given as
\begin{eqnarray}
&&
\xi=-\left({1\over 3}\ln H_2+{1\over 6}\ln H_5\right)
\nonumber \\
&&
\eta={1\over 6}\ln H_2+{1\over 3}\ln H_5
\nonumber \\
&&
\zeta_{2(\cdots 5)}={1\over 6}\left(\ln H_2-\ln H_5\right)
\nonumber \\
&&
\zeta_{6}=-{1\over 3}\left(\ln H_2-\ln H_5\right)
\,.
\end{eqnarray}
The index $2(\cdots 5)$ denotes
that it is either 2, 3, 4, or 5.
Hence the metric 
is given by 
\begin{eqnarray}
ds^2=H_2^{1/3}H_5^{2/3}\left[2(H_2H_5)^{-1}
du\left(dv+fdu+\frac{\cal A}{\sqrt{2}}\right)
+\sum_{\alpha=2}^5H_5^{-1}dy_\alpha^2
+H_2^{-1}dy_6^2+\sum_{i=1}^4dx_i^2\right]
\label{metric:M2-M5}
\,.
\end{eqnarray}

The non-trivial components of
field strength $F_{\hat A\hat B\hat C\hat D}$
are given by
\begin{eqnarray}
&&F_{\hat j\hat u\hat v\hat y_6}=
-H_2^{-1/6}H_5^{-1/3}\frac{\partial_jH_2}{H_2}\\
&&F_{\hat i\hat j\hat u\hat y_6}= 
-\frac{1}{\sqrt{2}}H_2^{-2/3}H_5^{-5/6}\left(
{\cal F}_{\hat i\hat j}^{(2)}+{}^*{\cal F}_{\hat i\hat j}^{(5)}
\right)\\
&&F_{\hat k\hat l\hat m\hat y_6}=\epsilon^{jklm}
H_2^{-1/6}H_5^{-1/3}\frac{\partial_jH_5}{H_5}
\,,
\end{eqnarray}
where 
${\cal F}_{\hat i\hat j}^{(A)}$ and those duals 
${}^*{\cal F}^{\hat i\hat j}_{(A)}$ are given by
\begin{eqnarray}
&&{\cal F}_{\hat i\hat j}^{(A)}\equiv
-2H_A\left(
\partial_{[i}B_{j]}^{(A)}-{\cal A}_{[i}\partial_{j]}
E_A\right)=-2H_A\left(
\partial_{[i}B_{j]}^{(A)}+{1\over H_A^2}{\cal A}_{[i}\partial_{j]}
H_A\right)
\\
&&
{}^*{\cal F}^{\hat i\hat j}_{(A)}\equiv
{1\over 2}\epsilon^{ijkl}{\cal F}_{kl}^{(A)}
\,.
\end{eqnarray}
If we have two charged branes (Case (1) in \S II:
$B_i^{(A)}=-{\cal A}_i/H_A$), 
each ${\cal F}_{\hat i\hat j}^{(A)}$ 
coincides with ${\cal F}_{\hat i\hat j}$.
For neutral branes with currents (Case (2) in \S II),
we find ${\cal F}_{\hat i\hat j}^{(A)}=\lambda_A {\cal F}_{\hat i\hat j}$.

Using Eq. (\ref{eqn:potential}) and the above explicit expression for
$F_{\hat A\hat B\hat C\hat D}$, 
we obtain the Killing equations (\ref{eqn:Killingspinor}) as 
\begin{eqnarray}
\delta\psi_{\hat u}
&=&
\frac{1}{6}H_2^{-1/6}H_5^{-1/3}
\left[\frac{\partial_jH_2}{H_2}\gamma^{\hat j\hat v}
(1-\gamma^{\hat u\hat v\hat y_6})
+\frac{1}{2}\frac{\partial_jH_5}{H_5}\gamma^{\hat j\hat v}
(1-\gamma^{\hat u\hat v\hat y_2\cdots\hat y_5})\right]
\epsilon
\nonumber\\
&&
-\frac{1}{4\sqrt{2}}H_2^{-2/3}H_5^{-5/6}
\left[\frac{1}{2}{\cal F}_{\hat i\hat j}\gamma^{\hat i\hat j}
-\frac{1}{3}\left(
{\cal F}_{\hat i\hat j}^{(2)}+{}^*{\cal F}_{\hat i\hat j}^{(5)}\right)
\gamma^{\hat i\hat j\hat y_6}
-\frac{1}{6}\left({\cal F}_{\hat i\hat j}^{(2)}+{}^*{\cal F}_{\hat i\hat j}^{(5)}\right)
{\gamma}^{\hat i\hat j\hat u\hat v\hat y_6}
\right]
\epsilon\nonumber\\
&&
-\frac{1}{2}H_2^{-1/6}H_5^{-1/3}\partial^jf\gamma^{\hat j\hat u}\epsilon
\nonumber\\
&=&
\frac{1}{6}H_2^{-1/6}H_5^{-1/3}
\left[\frac{\partial_jH_2}{H_2}\gamma^{\hat j\hat v}
(1-\gamma^{\hat u\hat v\hat y_6})
+\frac{1}{2}\frac{\partial_jH_5}{H_5}\gamma^{\hat j\hat v}
(1-\gamma^{\hat u\hat v\hat y_2\cdots\hat y_5})\right]
\epsilon
\nonumber\\
&&
-\frac{1}{4\sqrt{2}}H_2^{-2/3}H_5^{-5/6}
\left[\left\{\frac{1}{2}{\cal F}_{\hat i\hat j}
+{1\over 3}\left(
{}^*{\cal F}_{\hat i\hat j}^{(2)}+{\cal F}_{\hat i\hat j}^{(5)}\right)
-{1\over 6}\left(
{\cal F}_{\hat i\hat j}^{(2)}+{}^*{\cal F}_{\hat i\hat j}^{(5)}\right)
\right\}
\gamma^{\hat i\hat j}
\right.
\nonumber\\
&&
\left.
-\frac{1}{3}\left(
{}^*{\cal F}_{\hat i\hat j}^{(2)}+{\cal F}_{\hat i\hat j}^{(5)}\right)
\gamma^{\hat i\hat j}
(1-\gamma^{\hat u\hat v\hat y_2\cdots\hat y_5})
+\frac{1}{6}\left(
{\cal F}_{\hat i\hat j}^{(2)}+{}^*{\cal F}_{\hat i\hat j}^{(5)}\right)
\gamma^{\hat i\hat j}(
1-\gamma^{\hat u\hat v\hat y_6})
\right]
\epsilon
\nonumber\\
&&
-\frac{1}{2}H_2^{-1/6}H_5^{-1/3}\partial^jf\gamma^{\hat j\hat u}
\epsilon 
\,,
\end{eqnarray}

\begin{eqnarray}
\delta\psi_{\hat v}&=&\frac{1}{6}H_2^{-1/6}H_5^{-1/3}
\left[\frac{\partial_jH_2}{H_2}\gamma^{\hat j\hat u}
(1-\gamma^{\hat u\hat v\hat y_6})
 +
\frac{\partial_jH_5}{2H_5}\gamma^{\hat j\hat u}
(1-
\gamma^{\hat u\hat v\hat y_2\cdots\hat y_5})\right]
\epsilon
\,,
\end{eqnarray}

\begin{eqnarray}
\delta\psi_{\hat i}&=&
H_2^{-1/6}H_5^{-1/3}\partial_i\zeta
\nonumber \\
&&
+
\frac{1}{\sqrt{2}}H_2^{-2/3}H_5^{-5/6}
\left[
{1\over 4}
{\cal F}_{\hat i \hat j}\gamma^{\hat j\hat u}
+{1\over 6}\left({\cal F}_{\hat i\hat j}^{(2)}+{}^*{\cal F}_{\hat i\hat j}^{(5)}\right)
\gamma^{\hat j\hat u \hat y_6}
-{1\over 24}\left({\cal F}_{\hat j\hat k}^{(2)}+{}^*{\cal F}_{\hat j\hat k}^{(5)}\right)
\gamma^{\hat i\hat j\hat k\hat u \hat y_6}
\right] 
\epsilon
\nonumber\\
&&+\frac{1}{6}H_2^{-1/6}H_5^{-1/3}\left[\frac{\partial_jH_2}{2H_2}
\gamma^{\hat i\hat j}(1-\gamma^{\hat u\hat v\hat y_6})+
\frac{\partial^jH_5}{H_5}\gamma^{\hat i\hat j}
(1-\gamma^{\hat u\hat v\hat y_2\cdots\hat y_5})\right]
\epsilon
\nonumber\\
&&+\frac{1}{6}H_2^{-1/6}H_5^{-1/3}
\left[
\frac{\partial_iH_2}{H_2}
\gamma^{\hat u\hat v\hat y_6}
+ \frac{\partial_iH_5}{2H_5}
\gamma^{\hat u\hat v\hat y_2\cdots\hat y_5}
\right]
\epsilon 
\nonumber \\
&=&H_2^{-1/6}H_5^{-1/3}\left(\partial_i
+{\partial_i H_2\over 6H_2}+{\partial_i H_5\over 12H_5}\right)
\epsilon
\nonumber \\
&&
+
\frac{1}{\sqrt{2}}H_2^{-2/3}H_5^{-5/6}
\left[
\left\{
{1\over 4}
{\cal F}_{\hat i \hat j}
-{1\over 6}\left({\cal F}_{\hat i\hat j}^{(2)}+{}^*{\cal F}_{\hat i\hat j}^{(5)}\right)-{1\over 12}\left(
{}^*{\cal F}_{\hat i\hat j}^{(2)}+{\cal F}_{\hat i\hat j}^{(5)}
\right)
\right\}
\gamma^{\hat j\hat u}
\right.
\nonumber\\
&&
\left.
+{1\over 6}\left({\cal F}_{\hat i\hat j}^{(2)}+{}^*{\cal F}_{\hat i\hat j}^{(5)}\right)\gamma^{\hat j\hat u}(1-
\gamma^{\hat u \hat v\hat y_6})
+{1\over 12}\left({}^*{\cal F}_{\hat i\hat j}^{(2)}+{\cal F}_{\hat i\hat j}^{(5)}\right)\gamma^{\hat j\hat u}(1-
\gamma^{\hat u\hat v \hat y_2\cdots \hat y_5})
\right] 
\epsilon
\nonumber\\
&&+\frac{1}{6}H_2^{-1/6}H_5^{-1/3}\left[\frac{\partial_jH_2}{2H_2}
\gamma^{\hat i\hat j}(1-\gamma^{\hat u\hat v\hat y_6})+
\frac{\partial^jH_5}{H_5}\gamma^{\hat i\hat j}
(1-\gamma^{\hat u\hat v\hat y_2\cdots\hat y_5})\right]
\epsilon
\nonumber\\
&&-\frac{1}{6}H_2^{-1/6}H_5^{-1/3}
\left[
\frac{\partial_iH_2}{H_2}(1-
\gamma^{\hat u\hat v\hat y_6})
+ \frac{\partial_iH_5}{2H_5}
(1-\gamma^{\hat u\hat v\hat y_2\cdots\hat y_5})
\right]
\epsilon 
\,,
\end{eqnarray}

\begin{eqnarray}
\delta\psi_{\hat y_{2(\cdots5)}}
&=&-\frac{1}{12}H_2^{-1/6}H_5^{-1/3}\left[
\frac{\partial_jH_2}{H_2}\gamma^{\hat j\hat y_{2(\cdots5)}}
(1-\gamma^{\hat u\hat v\hat y_6})
-\frac{\partial_jH_5}{H_5}\gamma^{\hat j\hat y_{2(\cdots5)}}
(1-\gamma^{\hat u\hat v\hat y_2\cdots\hat y_5})\right]
\epsilon
\nonumber \\
&&
+{1\over 24\sqrt{2}}H_2^{-2/3}H_5^{-5/6}
\left(
{\cal F}^{(2)}_{\hat i\hat j}+{}^*{\cal F}^{(5)}_{\hat i\hat j}\right)
\gamma^{\hat i\hat j\hat u\hat y_{2(\cdots5)}\hat y_6}
\epsilon
\,,
\end{eqnarray}

\begin{eqnarray}
\delta\psi_{\hat y_6}
&=&\frac{1}{6}H_2^{-1/6}H_5^{-1/3}\left[
\frac{\partial_jH_2}{H_2}\gamma^{\hat j\hat y_6}
(1-\gamma^{\hat u\hat v\hat y_6})
-\frac{\partial_jH_5}{H_5}\gamma^{\hat j\hat y_6}
(1-\gamma^{\hat u\hat v\hat y_2\cdots\hat y_5})\right]
\epsilon
\nonumber \\
&&
-{1\over 12\sqrt{2}}H_2^{-2/3}H_5^{-5/6}\left(
{\cal F}^{(2)}_{\hat i\hat j}+{}^*{\cal F}^{(5)}_{\hat i\hat j}\right)
\gamma^{\hat i\hat j\hat u}
\epsilon
\,.
\end{eqnarray}

Most parts of the above equations vanish  if we impose
 the following condition for the Killing spinor $\epsilon$: 
\begin{eqnarray}
(1-\gamma^{\hat u\hat v\hat y_6})\epsilon=0,~
(1-\gamma^{\hat u\hat v\hat y_2\cdots\hat y_5})\epsilon=0,~
\gamma^{\hat{u}}\epsilon=0
\,.
\end{eqnarray}
These conditions can be rewritten as
\begin{eqnarray}
(1+\gamma^{\hat 0\hat y_1 \hat y_6})\epsilon=0,~~~
(1+\gamma^{\hat 0\hat y_1  \cdots\hat y_5})\epsilon=0,~~~
(1+\gamma^{\hat 0\hat y_1 })\epsilon=0
\,.
\label{cond:super}
\end{eqnarray}

However, two terms remain.
One term is
\begin{eqnarray}
\left(\partial_i
+{\partial_i H_2\over 6H_2}+{\partial_i H_5\over 12H_5}\right)
\epsilon
\,,
\label{condition1}
\end{eqnarray}
and the other term is
\begin{eqnarray}
\left\{\frac{1}{2}{\cal F}_{\hat i\hat j}
+{1\over 3}\left(
{}^*{\cal F}_{\hat i\hat j}^{(2)}+{\cal F}_{\hat i\hat j}^{(5)}\right)
-{1\over 6}\left(
{\cal F}_{\hat i\hat j}^{(2)}+{}^*{\cal F}_{\hat i\hat j}^{(5)}\right)
\right\}
\gamma^{\hat i\hat j}
\epsilon
\label{condition2}
\,.
\end{eqnarray}

The former term (\ref{condition1})
vanishes if  $\epsilon$
is described as
\begin{eqnarray}
\epsilon=H_2^{-1/6}H_5^{-1/12}\epsilon_0
\,,
\label{condition2}
\end{eqnarray}
where $\epsilon_0$ is a constant spinor. 
The latter term also vanishes if ${\cal F}^{(A)}_{ij}
\propto {\cal F}_{ij}$ ($A=2,5$) and
${\cal F}_{ij}$ is self-dual 
(${}^*{\cal F}_{ij}={\cal F}_{ij}$).
In fact, this term is proportional to
\begin{eqnarray}
{\cal F}_{\hat i\hat j}
\gamma^{\hat i\hat j}\epsilon
\,,
\end{eqnarray}
which vanishes for the self-dual field ${\cal F}_{ij}$
as shown below.
From Eqs. (\ref{cond:super}), we have
\begin{eqnarray}
\gamma^{\hat 0\hat y_1 }\epsilon=-\epsilon\,,~~~
\gamma^{\hat y_6}\epsilon=\epsilon\,,~~~
\gamma^{\hat y_2  \cdots\hat y_5}\epsilon=\epsilon
\,.
\end{eqnarray}
We also assume that 
\begin{eqnarray}
\gamma^{\hat 0 y_1 \cdots\hat y_6\hat x_1\hat x_2\hat x_3\hat x_4 }
\epsilon=-\epsilon
\,,
\end{eqnarray}
which corresponds to the chiral state.
Then, we find 
\begin{eqnarray}
\gamma^{\hat x_1\hat x_2\hat x_3\hat x_4 }
\epsilon=\epsilon
\,.
\end{eqnarray}
This equation is rewritten as
\begin{eqnarray}
\gamma^{\hat i\hat j}
\epsilon=-{1\over 2} \epsilon_{\hat i\hat j\hat k\hat l}\gamma^{\hat k\hat l}
\epsilon
\,.
\end{eqnarray}
We then obtain
\begin{eqnarray}
{\cal F}_{ij}\gamma^{\hat i\hat j}
\epsilon
=-{1\over 2} {\cal F}_{ij}\epsilon_{\hat i\hat j\hat k\hat l}
\gamma^{\hat k\hat l}
\epsilon
=-{}^*{\cal F}_{kl}\gamma^{\hat k\hat l}
\epsilon\,,
=-{\cal F}_{kl}\gamma^{\hat k\hat l}
\epsilon
\,.
\end{eqnarray}
The last equality is found by the self-duality of ${\cal F}_{ij}$.
This equation yields 
\begin{eqnarray}
{\cal F}_{ij}\gamma^{\hat i\hat j}
\epsilon=0
\,.
\end{eqnarray}
In the case of the BMPV type solution discussed in \S IV B,
this self-dual condition gives the relation between $J_\phi$ and
$J_\psi$, that is, $J_\phi=-J_\psi=J$.

The condition of ${\cal F}^{(A)}_{ij}\propto
{\cal F}_{ij}$ leads either to 
Case (1):two charged branes, or Case (2): neutral branes with currents
discussed in \S II.
We conclude that
 the solutions discussed in this paper preserve $1/8$ supersymmetry
if ${\cal F}_{ij}$ is self-dual \cite{bertolini}.

We also expect that the Kalza-Klein compactification into five-dimensional
spacetime does not break any 
supersymmetry, because all coordinates to
be compactified are cyclic. Thus the five dimensional solution
obtained here is a 
BPS state if ${\cal F}_{ij}$ is self-dual.

\section{``black" brane solutions in M-theory
 by use of different coordinate systems}

We may adopt  several different curvilinear 
coordinates for the following reason.
If we have a complete set of solutions, any solution can be described by
some linear combination of these solutions.
Therefore, we may not need to introduce another coordinate system.
However, if we use some coordinate system, we may be able to
describe some interesting solution by a few  multipole 
moments. In order to describe it
in the different coordinate system, one may need an infinite
series of eigenfunctions.
For example, a black ring solution may need 
an infinite sum
to describe it  in the Cartesian coordinate system.
But if we use an appropriate coordinate system, 
the expression could be much simpler.
Therefore, 
in this Appendix, we discuss two other curvilinear coordinates;
hyperelliptical and hyperpolorical coordinate systems.

We start with the metric form of 
the four-dimensional Euclidian space
written
by some orthonormal curvilinear coordinates,
i.e.,
\begin{eqnarray}
ds_{\mathbb{E}^4}^2=h_{\xi\xi}d\xi^2+h_{\eta\eta}d\eta^2
+h_{\phi\phi}d\phi^2+h_{\psi\psi}d\psi^2
\,.
\label{metric_form}
\end{eqnarray}
We assume that there are two rotation symmetries, 
as discussed in the text (\S IV).
Hence $h_{ij}$ depends only on two coordinates: $\xi$ and $\eta$.

Eqs. (\ref{eqn:H}), (\ref{eqn:A_phi}), (\ref{eqn:A_psi}) 
and (\ref{eqn:f}) 
are explicitly written as
\begin{eqnarray}
\partial_\xi\left(\sqrt{h_{\eta\eta}h_{\phi\phi}h_{\psi\psi}\over h_{\xi\xi}}
\partial_\xi H_A \right)+
\partial_\eta\left(\sqrt{h_{\xi\xi}h_{\phi\phi}h_{\psi\psi}\over h_{\eta\eta}}
\partial_\eta H_A \right)=0 ~~~(A=2,5)
\,,
\label{eqn:H1}
\end{eqnarray}
\begin{eqnarray}
&&
\partial_\xi\left(\sqrt{h_{\eta\eta}h_{\psi\psi}\over h_{\xi\xi}h_{\phi\phi}}
\partial_\xi {\cal A}_\phi \right)+
\partial_\eta\left(\sqrt{h_{\xi\xi}h_{\psi\psi}\over h_{\eta\eta}h_{\phi\phi}}
\partial_\eta {\cal A}_\phi \right)=0
\label{eqn:Aphi1}\\
&&
\partial_\xi\left(\sqrt{h_{\eta\eta}h_{\phi\phi}\over h_{\xi\xi}h_{\psi\psi}}
\partial_\xi {\cal A}_\psi \right)+
\partial_\eta\left(\sqrt{h_{\xi\xi}h_{\phi\phi}\over h_{\eta\eta}h_{\psi\psi}}
\partial_\eta {\cal A}_\psi \right)=0
\label{eqn:Apsi1}
\,,
\end{eqnarray}
\begin{eqnarray}
&&\partial_\xi\left(\sqrt{h_{\eta\eta}h_{\phi\phi}h_{\psi\psi}\over h_{\xi\xi}}
\partial_\xi f \right)+
\partial_\eta\left(\sqrt{h_{\xi\xi}h_{\phi\phi}h_{\psi\psi}\over h_{\eta\eta}}
\partial_\eta f \right)\nonumber \\
&&={\beta \over 4H_2 H_5}
\left[
\sqrt{h_{\eta\eta}h_{\psi\psi}\over h_{\xi\xi}h_{\phi\phi}
}(\partial_\xi{\cal A}_\phi)^2+
\sqrt{h_{\xi\xi}h_{\psi\psi}\over 
h_{\eta\eta}h_{\phi\phi}}(\partial_\eta{\cal A}_\phi)^2+
\sqrt{h_{\eta\eta}h_{\phi\phi}\over h_{\xi\xi}h_{\psi\psi}}
(\partial_\xi{\cal A}_\psi)^2+
\sqrt{h_{\xi\xi}h_{\phi\phi}\over h_{\eta\eta}h_{\psi\psi}}
(\partial_\eta{\cal A}_\psi)^2
\right]
\,.
\label{eq:f1}
\end{eqnarray}

The ADM mass is determined from the asymptotic behaviors of  $H_A$ and $f$
as
\begin{eqnarray}
M_{\rm ADM}={\pi\over 4G_5}\left({\cal Q}_0+{\cal Q}_H^{(2)}
+{\cal Q}_H^{(5)}
\right)
\,,
\end{eqnarray}
where
\begin{eqnarray}
H_A&\rightarrow&1+{{\cal Q}_H^{(A)}\over r^2}\nonumber \\
f&\rightarrow&{{\cal Q}_0\over r^2}
\,,
\end{eqnarray}
as 
\begin{eqnarray}
r\equiv\left({\sum_{i=1}^4 x_i^2}\right)^{1/2}\rightarrow \infty
\,.
\end{eqnarray}
If the  horizon is given by $\xi=\xi_h$,
the area  $A_h$ is  
\begin{eqnarray}
 A_h&=4\pi^2 &\int_{\xi=\xi_h} d\eta \left(
 h_{\eta\eta}h_{\phi\phi}h_{\psi\psi}\right)^{1/2}
 \left[H_2H_5(1+f)-{1\over 8}\left(\frac{{\cal A}_\phi^2}{h_{\phi\phi}}
 +\frac{{\cal A}_\psi^2}{h_{\psi\psi}}\right)\right]^{1/2}
\,,
 \end{eqnarray}
which gives the entropy of a ``black" object.

\subsection{hyperelliptical coordinates}
First we adopt the hyperelliptical coordinates
$(\xi,\eta,\phi,\psi)$, which are defined by
the following transformation:
\begin{eqnarray}
x_1+ix_2=R\cosh \xi\cos \eta e^{i\phi},\quad 
x_3+ix_4=R\sinh \xi \sin \eta e^{i\psi}
\,,
\end{eqnarray}
where $R$ is a constant, $\xi \geq 0$, $0\leq \eta \leq \pi$, 
and $0\leq \phi, \psi \leq 2\pi$.

The line element is given by
\begin{eqnarray}
ds_{\mathbb{E}^4}^2=R^2\left[(\sinh^2\xi+\sin^2\eta)
(d\xi^2+d\eta^2)
+ \cosh^2\xi\cos^2\eta d\phi^2
+\sinh^2\xi\sin^2\eta d\psi^2 \right]
\,.
\end{eqnarray}

Eq. (\ref{eqn:H1}) is 
\begin{eqnarray}
\frac{1}{\sinh \xi\cosh \xi}\partial_\xi
(\sinh \xi\cosh \xi\partial_\xi H_A)+
\frac{1}{\sin \eta\cos \eta}\partial_\eta
(\sin \eta \cos \eta\partial_\eta H_A)=0
~~~(A=2,5)
\,.
\end{eqnarray}

Setting $H_A=h_A(\xi)j_A(\eta)$, we 
find two ordinary differential equations:
\begin{eqnarray}
&&\frac{1}{\sinh \xi\cosh \xi}\frac{d}{d\xi}\left(
\sinh \xi\cosh \xi\frac{dh_A}{d\xi}\right)-Mh_A=0,\\
&&\frac{1}{\sin \eta\cos \eta}\frac{d}{d\eta}\left(
\sin \eta\cos \eta\frac{dj_A}{d\eta}\right)+Mj_A=0
\,,
\end{eqnarray}
where $M$ is a separation constant.
Using new variables $\rho=\cosh  2\xi $ and $\mu=\cos  2\eta $,
these equations are rewritten by the Legendre equation as 
\begin{eqnarray}
&&\frac{d}{d\rho}\left((\rho^2-1)\frac{dh_A}{d\rho}\right)
-\frac{M}{4}h_A=0, \\
&&\frac{d}{d\mu}\left((1-\mu^2)\frac{dj_A}{d\mu}\right)
+\frac{M}{4}j_A=0
\,.
\end{eqnarray}
The regularity condition on the symmetric axis
gives $M=4\ell(\ell+1)\quad 
(\ell=0,1,2,\cdots)$.
A general solution for $H_A$ is then 
\begin{eqnarray}
H_A=\sum_{\ell=0}^\infty
\left[g_\ell^{(A)} P_\ell(\cosh 2\xi )+h_\ell^{(A)} Q_\ell(\cosh 2\xi)\right] 
P_\ell(\cos 2\eta )
\,,  
\label{sol:H_elliptical}
\end{eqnarray}
where $Q_\ell(z)$ is the second kind Legendre function, and 
$g_\ell^{(A)} $ and $h_\ell^{(A)}$ are arbitrary constants.

>From the condition of asymptotically flatness
($H_A\rightarrow 1$ as $r\rightarrow \infty$), $H_A$ is given by  
\begin{eqnarray}
H_A=1+\sum_{\ell=0}^\infty h_\ell^{(A)}
 Q_\ell(\cosh 2\xi)P_\ell (\cos 2\eta)
~~~(A=2,5)
\,,
\end{eqnarray}
because $Q_\ell(z)$  vanishes
at $z=\infty$.
The explicit form for $\ell=0,1,2$ is
as follows:
\begin{eqnarray}
Q_0(z)&=&-{1\over 2}\ln\left({z+1\over z-1}\right)
\nonumber \\
Q_1(z)&=&{z\over 2}\ln\left({z+1\over z-1}\right)-1
\nonumber \\
Q_2(z)&=&{1\over 4}\left(3z^2-1\right)
\ln\left({z+1\over z-1}\right)-{3z\over 2}
\,.
\end{eqnarray}

The solution of $H_A$ with the lowest moment ($\ell=0$)
is
\begin{eqnarray}
H_A=1+h_0^{(A)} \ln\left(\tanh \xi\right)
\,.
\end{eqnarray}
If we define a charge ${\cal Q}_H^{(A)}$
 by the asymptotic behavior
of $H_A$ as $H_A\rightarrow 1+{\cal Q}_H^{(A)}/r^2$,
we find that
\begin{eqnarray}
h_0^{(A)} =-{2{\cal Q}_H^{(A)}\over R^2}
\,,
\end{eqnarray}
because $\ln (\tanh \xi)\sim -2/e^{2\xi}\approx -R^2/(2r^2)$.

Now we solve Eqs.  (\ref{eqn:Aphi1}) and (\ref{eqn:Apsi1}), which are 
\begin{eqnarray}
&&\coth \xi\partial_\xi(\tanh \xi\partial_\xi{\cal A}_\phi)
+\cot \eta\partial_\eta(\tan \eta\partial_\eta{\cal A}_\phi)=0,\\
&&\tanh \xi\partial_\xi(\coth \xi\partial_\xi{\cal A}_\psi)
+\tan \eta\partial_\eta(\cot \eta\partial_\eta{\cal A}_\psi)=0
\,.
\end{eqnarray}

Setting ${\cal A}_\phi=a_\phi (\xi)b_\phi(\eta)$ and 
${\cal A}_\psi=a_\psi(\xi)b_\psi(\eta)$, we obtain the 
following ordinary differential equations: 
\begin{eqnarray}
&&\frac{d^2a_\phi}{d\rho^2}
+\frac{1}{\rho-1}\frac{da_\phi}{d\rho}
-\frac{K}{\rho^2-1}a_\phi=0,
\label{Aphi_u}
\\
&&\frac{d^2b_\phi}{d\mu^2}
-\frac{1}{1-\mu}\frac{db_\phi}{d\mu}
+\frac{K}{1-\mu^2}b_\phi=0,
\label{Aphi_v}
\\
&&\frac{d^2a_\psi}{d\rho^2}
+\frac{1}{\rho+1}\frac{da_\psi}{d\rho}
-\frac{L}{\rho^2-1}a_\psi=0,
\label{Apsi_u}
\\
&&\frac{d^2b_\psi}{d\mu^2}
+\frac{1}{1+\mu}\frac{db_\psi}{d\mu}
+\frac{L}{1-\mu^2}b_\psi=0
\label{Apsi_v}
\,,
\end{eqnarray}
where $\rho=\cosh 2\xi $ and $ \mu=\cos 2 \eta$, 
and $K$ and $L$ are separation constants. 

Eqs. (\ref{Aphi_v}) and (\ref{Apsi_v}) are the same as Eqs 
(\ref{eqn:Aphi_sphere2}) and (\ref{eqn:Apsi_sphere2}).
Then we obtain the angular solutions 
by hypergeometric functions as
$b_\phi=F\left(-m,
m,1,(1-\mu)/2\right)$ and
 $b_\psi=F\left(-n,
n,1,(1+\mu)/2\right)$.
We have set $K=m^2$ and $L=n^2$ ($m,n=1,2,\ldots$)
from regularity conditions on the symmetric axis. 
The solutions for Eqs. (\ref{Aphi_u}) and (\ref{Apsi_u}) are also 
given by the hypergeometric functions.
Imposing the asymptotically flatness condition at
 infinity ($\xi\rightarrow
\infty$), we find the following solutions: 
\begin{eqnarray}
&&{\cal A}_\phi=\sum_{m=1}^\infty b_m^{(\phi)}\sinh^{-2m}\xi
F\left(m, m, 1+2m, -\sinh^{-2}\xi\right)F(-m, m, 1, \sin^2\eta)
\\
&&{\cal A}_\psi=\sum_{n=1}^\infty b_n^{(\psi)}\cosh^{-2n}\xi
F\left(n, n, 1+2n, \cosh^{-2}\xi\right)F(-n, n, 1, \cos^2\eta)
\,.
\end{eqnarray}

Here we show some hypergeometric functions, which we use later,
 explicitly:
\begin{eqnarray}
F(1, 1, 3, z)&=&{2\over z^2}\left[
z+(1-z)\ln (1-z) \right]
\nonumber \\
F(1, 2, 3, z)&=&-{2\over z^2}
\left[z+\ln (1-z) \right]
\,.
\end{eqnarray}

If $\beta=0$ (two charged branes or neutral branes with
appropriately chosen current strength $\lambda_{A''}$),
the solution of $f$ is given by a harmonic function, which is
\begin{eqnarray}
f=\sum_{\ell=0}^\infty c_\ell
 Q_\ell(\cosh 2\xi)P_\ell (\cos 2\eta)
\,,
\end{eqnarray}
where 
\begin{eqnarray}
c_\ell=-{2{\cal Q}_\ell\over R^2}
\,.
\end{eqnarray}

The lowest moment solution ($\ell=0, m=n=1$) in this case
is
\begin{eqnarray}
&&
H_A=1-{2{\cal Q}_H^{(A)}\over R^2} \ln\left(\tanh \xi\right)~~~(A=2,5)\\
&&
f=-{2{\cal Q}_0\over R^2} \ln\left(\tanh \xi\right)\\
&&{\cal A}_\phi
=-2J_1^{(\phi)}\left[1+2\cosh^2\xi ~\ln (\tanh \xi)
\right]\cos^2\eta
\\
&&{\cal A}_\psi
=2 J_1^{(\psi)}\left[1+2\sinh^2\xi ~\ln (\tanh \xi)
\right]\sin^2\eta
\,,
\end{eqnarray}
where ${\cal Q}_H^{(2)}, {\cal Q}_H^{(5)}, {\cal Q}_0$ are charges
and $J_1^{(\phi)},J_1^{(\psi)}$ are angular momentum.
We can show that this spacetime is supersymmetric
if $J_1^{(\phi)}=-J_1^{(\psi)}$, which is the same condition
as that for the BMPV black hole solution.

The ADM mass of this object is
\begin{eqnarray}
M_{\rm ADM}&=&\frac{\pi}{4G_5}\left({\cal Q}_0
+{\cal Q}_H^{(2)}+{\cal Q}_H^{(5)}
\right)
\,.
\end{eqnarray}

$H_A$ and $(1+f)$ diverge at $\xi=0$,
which may correspond to the horizon.
Calculating the Kretschmann curvature invariant,
we show that it is a naked singularity.
Therefore, this solution does not provide a black hole spacetime,
but instead, describes the spacetime of a rotating singular disk.

In the case of $\beta\neq 0$,
in order to obtain the solution for $f$,
we have to expand $f$ and the source term $\tilde{\cal S}$
by the Legendre function $P_\ell(\cos 2\eta)$, just as in the case
 of the previous
hyperspherical coordinates
(Eqs. (\ref{expand_f_Legendre}) and (\ref{expand_S_Legendre}).
$\tilde{\cal S}$ is defined by
\begin{eqnarray}
\tilde{\cal S}(\xi,\eta)&\equiv
&R^2(\sinh^2 \xi+\sin^2 \eta)~{\cal S}(\xi,\eta)
=
{\beta R^2(\sinh^2 \xi+\sin^2 \eta)
\over 8H_2 H_5}{\cal F}_{ij}{\cal F}^{ij}
\nonumber \\
&=&{\beta\over 4R^2H_2 H_5}
\left[\frac{1}{\cosh^2\xi\cos^2\eta}\left((\partial_\xi{\cal A}_\phi)^2
+(\partial_\eta{\cal A}_\phi)^2\right)+\frac{1}{\sinh^2\xi\sin^2\eta}
\left((\partial_\xi{\cal A}_\psi)^2+(\partial_\eta
{\cal A}_\psi)^2\right)\right]
\,.
\end{eqnarray}

We then have
\begin{eqnarray}
\frac{d}{d\rho}\left((\rho^2-1)\frac{df_\ell}{d\rho}\right)
-\ell(\ell+1) f_\ell={\tilde{\cal S}_\ell\over 4}
\,,
\label{eq:f_ell}
\end{eqnarray}
where $\rho=\cosh 2\xi$.

Let us show one concrete example, which is 
the lowest moment solution ($m=n=1$).
Setting $H_2=H_5=1$ (Case (2c) in Table II)
and
\begin{eqnarray}
&&{\cal A}_\phi
=-2J_1^{(\phi)}\left[1+2\cosh^2\xi ~\ln (\tanh \xi)
\right]\cos^2\eta
\\
&&{\cal A}_\psi
=2 J_1^{(\psi)}\left[1+2\sinh^2\xi ~\ln (\tanh \xi)
\right]\sin^2\eta
\,,
\end{eqnarray}
we find
\begin{eqnarray}
\tilde{\cal S}_0(\rho)&=&{8\beta J^2\over R^2}\left[
{2\rho\over \rho^2-1}+2\ln\left({\rho-1\over \rho+1}\right)
+{\rho\over 2}\left(\ln\left({\rho-1\over \rho+1}\right)\right)^2\right]
\\
\tilde{\cal S}_1(\rho)&=&{8\beta J^2\over R^2}
\left[
{2\over \rho^2-1}-{1\over 2}
\left(\ln\left({\rho-1\over \rho+1}\right)\right)^2\right]
\,,
\end{eqnarray}
where 
$2J^2=(J_1^{(\phi)})^2+(J_1^{(\psi)})^2$.
Integrating Eq. (\ref{eq:f_ell}), we find the exact solution
as 
\begin{eqnarray}
f(\xi,\eta)=f_0(\xi)+f_1(\xi)P_1(\cos 2\eta)
\,,
\end{eqnarray}
 with
\begin{eqnarray}
f_0(\xi)&=&-{2{\cal Q}_0\over R^2}\ln (\tanh \xi)
+{2\beta J^2\over R^2}\cosh 2\xi
\left[\ln (\tanh \xi)\right]^2
\\
f_1(\xi)&=&{2{\cal Q}_1\over R^2}\left[1+\cosh 2\xi \ln (\tanh \xi)\right]
-{\beta J^2\over 2R^2}\left[\ln (\tanh \xi)\right]^2
\,.
\end{eqnarray}

The ADM mass is 
\begin{eqnarray}
M_{\rm ADM}&=&\frac{\pi}{4G_5}{\cal Q}_0
\,.
\end{eqnarray}

Although this is an exact solution, 
it is very complicated.
Unless $\beta=0$, the horizon, even if it exists,
is not described by a surface of $\xi$= constant.

Note that although this solution is very complicated,
it is still supersymmetric if $J_1^{(\phi)}=-J_1^{(\psi)}$.

\subsection{hyperpolorical coordinates}

Our next example is the hyperpolorical coordinates 
$(\xi, \eta, \phi, \psi)$, which are defined by the transformation 
\begin{eqnarray}
x_1+ix_2=\frac{R\sinh \xi}{\cosh \xi-\cos \eta}e^{i\psi},\quad 
x_3+ix_4=\frac{R\sin \eta}{\cosh \xi-\cos \eta}e^{i\phi}
\,,
\end{eqnarray}
where $\xi \geq 0$, $0\leq \eta \leq \pi$, 
and $0\leq \phi, \psi \leq 2\pi$.
This coordinates could be used to describe a ring topology.
In this case, the  infinity corresponds to
$\xi=0$, which also describes one of the symmetric axis.

The line element is given by
\begin{eqnarray}
ds_{\mathbb{E}^4}^2
=\frac{R^2}{(\cosh \xi-\cos \eta)^2}(d\xi^2+\sinh^2\xi d\psi^2+d\eta^2
+\sin^2\eta d\phi^2)
\,.
\end{eqnarray}

With this coordinate system, 
Eq.  (\ref{eqn:H1}) is written as
\begin{eqnarray}
\frac{1}{\sinh \xi}\partial_\xi\left(\frac{\sinh \xi}{(\cosh \xi-\cos \eta)^2}
\partial_\xi H_A\right)+\frac{1}{\sin \eta}\partial_\eta\left(
\frac{\sin \eta}{(\cosh \xi-\cos \eta)^2}\partial_\eta H_A\right)=0, 
\,.
\label{eq:hyperpolorical_H}
\end{eqnarray}

Using new variable $\tilde{H}_A$, which is defined by  $H_A(\xi,\eta)=
1+(\cosh \xi-\cos \eta)\tilde{H}_A(\xi,\eta)$,
Eq. (\ref{eq:hyperpolorical_H}) is rewritten as
\begin{eqnarray}
\partial_\xi^2\tilde{H}_A+\coth \xi \partial_\xi \tilde{H}_A
+\partial_\eta^2\tilde{H}_A+\cot \eta\partial_\eta \tilde{H}_A=0
\,.
\end{eqnarray}

Setting $\tilde{H}_A=\tilde{h}_A(\xi)\tilde{j}_A(\eta)$, 
we can separate the variables and 
find the following
two ordinary differential equations: 
\begin{eqnarray}
&&(\rho^2-1)\frac{d^2\tilde{h}_A}{d\rho^2}+2\rho
\frac{d\tilde{h}_A}{d\rho}-M\tilde{h}_A=0,\\
&&(1-\mu^2)\frac{d^2\tilde{j}_A}{d\mu^2}-2\mu\frac{d\tilde{j}_A}{d\mu}
+M\tilde{j}_A=0
\,,
\end{eqnarray}
where $\rho=\cosh \xi$ and $\mu=\cos \eta$, and $M$ 
is a separation constant. 

We find that the general solution is described by the Legendre functions as
\begin {eqnarray}
H_A=1+(\cosh \xi-\cos \eta)\sum_{\ell=0}^\infty\left[h_\ell^{(A)} 
P_\ell(\cosh \xi)+g_\ell^{(A)}  Q_\ell (\cosh \xi)\right] 
P_\ell (\cos \eta)
\,,
\end{eqnarray}
where the separation constant $M$ 
is given by an integer $\ell$ as $M=\ell(\ell+1)$ because of 
the regularity on the symmetric axis. 
$g_\ell^{(A)}$ and $h_\ell^{(A)}$ are arbitrary constants.

The asymptotically flatness condition
yields 
\begin{eqnarray}
H_A=1+(\cosh \xi-\cos \eta)\sum_{\ell=0}^\infty h_\ell^{(A)}  
P_\ell(\cosh \xi)P_\ell(\cos \eta)
\,.
\end{eqnarray}  
Since 
$r^2=R^2(\cosh \xi+\cos \eta)/(\cosh \xi-\cos \eta)$,
looking at the asymptotic behavior at infinity,
we find $(\cosh \xi-\cos \eta)\sim 2R^2/r^2$ as
$r\rightarrow \infty(\xi,\eta\rightarrow 0)$.
This gives a relation between the coefficient $h_0^{(A)}$ and 
charge ${\cal Q}_H$ as
\begin{eqnarray}
h_0^{(A)} ={{\cal Q}_H^{(A)} \over 2R^2}
\,.
\end{eqnarray}

Now we discuss Eqs. (\ref{eqn:Aphi1}) and (\ref{eqn:Apsi1}), which are 
\begin{eqnarray}
&&\frac{1}{\sinh \xi}\partial_\xi(\sinh \xi\partial _\xi{\cal A}_\phi)
+\sin \eta\partial_\eta
\left(\frac{1}{\sin \eta}\partial_\eta{\cal A}_\phi\right)=0\\
&&\sinh \xi\partial_\xi
\left(\frac{1}{\sinh \xi}\partial _\xi{\cal A}_\psi\right)
+\frac{1}{\sin \eta}\partial_\eta(\sin \eta\partial_\eta{\cal A}_\psi)=0
\,.
\end{eqnarray}

Setting ${\cal A}_\phi=a_\phi(\xi)b_\phi(\eta)$ and 
${\cal A}_\psi=a_\psi(\xi)b_\psi(\eta)$, we obtain the following
ordinary differential equations: 
\begin{eqnarray}
&&\frac{d^2a_\phi}{d\rho^2}+\frac{2\rho}{\rho^2-1}
\frac{da_\phi}{d\rho}-\frac{K}{\rho^2-1}a_\phi=0,
\label{eq:a_phi}\\
&&\frac{d^2b_\phi}{d\mu^2}+\frac{K}{1-\mu^2}b_\phi=0,
\label{eq:b_phi}\\
&&\frac{d^2a_\psi}{d\rho^2}-\frac{L}{\rho^2-1}a_\psi=0,
\label{eq:a_psi}\\
&&\frac{d^2b_\psi}{d\mu^2}-\frac{2\mu}{1-\mu^2}
\frac{db_\psi}{d\mu}+\frac{L}{1-\mu^2}b_\psi=0
\,,
\label{eq:b_psi}
\end{eqnarray}
where $\rho=\cosh \xi$ and $\mu=\cos \eta$, and $K$ and $L$ are 
separation constants. 

The solutions for Eqs. (\ref{eq:b_phi}) and (\ref{eq:b_psi}) are given by 
the Legendre functions. We set $K=m(m+1)$ and $L=n(n+1)$
($m,n=1,2,\cdots$)
because of regularity conditions on the symmetric axis.

The asymptotically flatness condition
yields 
\begin{eqnarray}
&&{\cal A}_\phi=\sum_{m=1}^\infty  {b_m^{(\phi)}\over m+1}
P_{m}(\cosh \xi)
\left[\cos\eta ~P_m(\cos\eta)-P_{m-1}(\cos\eta)
\right]\\
&&{\cal A}_\psi=\sum_{n=1}^\infty {b_n^{(\psi)} \over n+1}
\left[
\cosh \xi ~P_n(\cosh\xi)-P_{n-1}(\cosh\xi)\right]
P_{n}(\cos \eta)
\,,
\end{eqnarray}
where $b_m^{(\phi)}$ and $b_n^{(\psi)} $ are arbitrary constants.

When $\beta=0$,
$f$ is given by 
the Legendre functions just as $H_A$, i.e.,\\
\begin{eqnarray}
f=(\cosh \xi-\cos \eta)\sum_{\ell=0}^\infty c_\ell 
P_\ell(\cosh \xi)P_\ell(\cos \eta)
\,,
\end{eqnarray}  
where $c_\ell$'s are  arbitrary constants.
For the lowest moment solution, 
we find
\begin{eqnarray}
H_A(\xi,\eta)&=&1+{{\cal Q}_H^{(A)}\over 2R^2}(\cosh\xi-\cos\eta),\\
f(\xi,\eta)&=&{{\cal Q}_0\over 2R^2}(\cosh\xi-\cos\eta),\\
{\cal A}_\phi
&=&J_1^{(\phi)}\cosh \xi \sin^2\eta\\
{\cal A}_\psi
&=&J_1^{(\psi)}\sinh^2 \xi
\cos \eta
\,.
\end{eqnarray}
The self-dual condition for supersymmetry implies
$J_1^{(\phi)}=-J_1^{(\psi)}$.

In this spacetime, there is no horizon, but rather,  a singularity
at $\xi=\infty$, which locates at a ring with a radius
$R$ in the flat 4D 
Euclidian space.
Then this describes the geometry of a ring singularity.

To solve the equation for $f$
in the case of $\beta\neq 0$, 
we again expand $f$ and the source term
$\tilde{\cal S}$ by the Legendre functions as
\begin{eqnarray}
f(\xi, \eta)&\equiv&(\rho-\mu)\sum_{\ell=0}^\infty  
\tilde{f}_\ell(\rho)P_\ell(\mu)
\\
\tilde{\cal S}(\xi, \eta)&\equiv&
{R^2\over (\cosh \xi-\cos \eta)^3}~{\cal S}(\xi, \eta)
~=~{\beta R^2\over 8H_2H_5(\cosh \xi-\cos \eta)^3}
{\cal F}_{ij}{\cal F}^{ij}
\nonumber \\
&=&{\beta(\cosh \xi-\cos \eta)\over 4R^2H_2H_5}\left[
{1\over \sin ^2\eta}\left((\partial_\xi {\cal A}_\phi)^2
+(\partial_\eta {\cal A}_\phi)^2
\right)
+{1\over \sinh ^2\xi}\left((\partial_\xi {\cal A}_\psi)^2
+(\partial_\eta {\cal A}_\psi)^2
\right)
\right]\nonumber \\
&=&\sum_{\ell=0}^\infty \tilde{\cal S}_\ell(\rho)
P_\ell(\mu)
\,,
\end{eqnarray}
where $\rho=\cosh\xi$ and $\mu=\cos\eta$.

We obtain the equation for $f_\ell(\rho)$ for each $\ell$ as
\begin{eqnarray}
(\rho^2-1) \frac{d^2 \tilde{f}_\ell}{d\rho^2}
+2\rho \frac{d\tilde{f}_\ell}{d\rho}
-\ell(\ell+1) \tilde{f}_\ell=\tilde{\cal S}_\ell(\rho)
\,,
\end{eqnarray}
where $\rho=\cosh\xi$.

Setting $H_2=H_5=1$ (Case (2c) in Table II) and
\begin{eqnarray}
&&{\cal A}_\phi
=J_1^{(\phi)}\cosh \xi \sin^2\eta\\
&&{\cal A}_\psi
=J_1^{(\psi)}\sinh^2 \xi
\cos \eta
\,,
\end{eqnarray}
we show the lowest moment solution here.
We have now
\begin{eqnarray}
\tilde{\cal S}_0&=&{\beta J^2\over 3R^2}\rho\left(3\rho^2-1\right)\nonumber \\
\tilde{\cal S}_1&=&-{\beta J^2\over 5R^2}\left(7\rho^2-1\right)\nonumber \\
\tilde{\cal S}_2&=&{\beta J^2\over 3R^2}\rho\left(3\rho^2+1\right)\nonumber \\
\tilde{\cal S}_3&=&-{\beta J^2\over 5R^2}\left(3\rho^2+1\right)
\,,
\end{eqnarray}
where $2J^2\equiv [(J_1^{(\phi)})^2+(J_1^{(\psi)})^2]$, 
and then find general solutions as
\begin{eqnarray}
\tilde{f}_0(\rho)&=&c_0P_0(\rho)+d_0 Q_0(\rho)
+{\beta J^2\over 24R^2}\left[2\rho(\rho^2+1)
+\ln\left(\rho-1\over \rho+1
\right)\right]\nonumber \\
\tilde{f}_1(\rho)&=&c_1P_1(\rho)+d_1 Q_1(\rho)
-{\beta J^2 \over 40R^2}\rho \left[14\rho 
+5\ln\left(\rho-1\over \rho+1
\right)\right]\nonumber \\
\tilde{f}_2(\rho)&=&c_2P_2(\rho)+d_2 Q_2(\rho)
-{\beta J^2\over 24R^2}\rho\left[
-2\rho(2\rho^2-1)-(3\rho^2-1)
\ln\left(\rho-1\over \rho+1
\right)\right]\nonumber \\
\tilde{f}_3(\rho)&=&c_3P_3(\rho)+d_3 Q_3(\rho)+
{\beta J^2\over 10R^2}\rho^2
\,.
\end{eqnarray}
Imposing the regularity ($\tilde{f}_0$ : finite, 
$\tilde{f}_\ell=0$ for $\ell 
\geq 1$)
 at infinity and on the axis ($\rho=1$),
we can fix the coefficients $c_\ell$ and $d_\ell$ $ (\ell=0,1,2,3)$
except for $c_0$.
We obtain
an exact solution as
\begin{eqnarray}
f(\xi,\eta)=(\cosh\xi-\cos\eta)\sum_{\ell=0}^{3}
\tilde{f}_\ell(\cosh\xi)P_\ell(\cos\eta)
\,,
\end{eqnarray}
with 
\begin{eqnarray}
\tilde{f}_0(\rho)&=&{\beta J^2\over 12R^2}(\rho-1)(\rho^2+\rho+2)
+{{\cal Q}_0\over 2R^2}
\nonumber \\
\tilde{f}_1(\rho)&=&-{\beta J^2\over 20R^2}(\rho-1)(7\rho+5)
\nonumber \\
\tilde{f}_2(\rho)&=&+{\beta J^2\over 12R^2}(\rho-1)(2\rho^2+5\rho+1)
\nonumber \\
\tilde{f}_3(\rho)&=&
-{\beta J^2\over 20R^2}\rho(\rho-1)(5\rho+3)
\,,
\end{eqnarray}
where ${\cal Q}_0$ is an arbitrary charge.
The ADM mass is given by
\begin{eqnarray}
M_{\rm ADM}&=&\frac{\pi}{4G_5} {\cal Q}_0
\,.
\end{eqnarray}

This exact solution is also very complicated, but supersymmetric
if $J_1^{(\phi)}=-J_1^{(\psi)}$.
The horizon, even if it exists,
is not described by a surface of $\xi$=constant.



\begin{thebibliography}{99}

\bibitem{Strominger_Vafa} A. Strominger and C. Vafa, 
Phys. Lett. B 379 (1996) 99, hep-th/9601029.
\bibitem{Gibbons} G.W. Gibbons, Nucl. Phys. B 207 (1982) 337.
\bibitem{Myers} R.C. Myers, Nucl. Phys. B 289 (1987) 701.
\bibitem{Gibbons_Maeda} G. W. Gibbons, and K. Maeda, 
Nucl. Phys. B 298 (1988) 741.
\bibitem{CMP} C.G. Callan, R.C. Myers, and M.J. Perry, 
Nucl. Phys. B 311 (1988) 673.
\bibitem{GHS} D. Garfinkle, G.T. Horowitz and A. Strominger, \\
Phys. Rev. D 43 (1991) 3140, Erratum-ibid. D 45 (1992) 3888. 
\bibitem{Horowitz_Strominger} G.T. Horowitz and A. Strominger, 
Nucl.Phys. B 360 (1991)197.
\bibitem{Tangherlini}F. R. Tangherlini, 
Nuovo Cim. 27 (1963) 636.
\bibitem{Myers_Perry} R.C. Myers and M.J. Perry, 
Ann. Phys. 172 (1986) 304.
\bibitem{GIS} G. W. Gibbons, D. Ida and T. Shiromizu, 
Phys. Rev. Lett. 89 (2002) 041101, hep-th/0206136.
\bibitem{SUSYunique} H. S. Reall,
Phys. Rev. D 68 (2003) 024024, hep-th/0211290.
\bibitem{Elvang_Emparan}H. Elvang, R. Emparan
JHEP 0311 (2003) 035, hep-th/0310008. 
\bibitem{Khuri_Myers}R. R. Khuri, and R. C. Myers, 
Fields Inst. Comm. 15 (1997) 273, hep-th/9512137.
\bibitem{Duff_Lu}
M.J. Duff and J.X. Lu, 
Nucl. Phys. B 416 (1994) 301, hep-th/9306052.
\bibitem{DLP}
M.J. Duff, J.X. Lu and C.N. Pope, 
Phys. Lett. B 382 (1996) 73, hep-th/9604052.
\bibitem{Callan_Maldacena}C. G. Callan, J. M. Maldacena, 
Nucl. Phys. B472 (1996) 591, hep-th/9602043.
\bibitem{Emparan_Reall}
R. Emparan and H.R. Reall, 
Phys. Rev. Lett. 88 (2000) 101101, hep-th/0110260.
\bibitem{Kallosh}R. Kallosh, Phys. Lett. B 282 (1992)
80, hep-th/9201029.
\bibitem{FKS} S. Ferrara, R. Kallosh, A. Strominger, 
Phys. Rev. D 52 (1995) 5412, hep-th/9508072.
\bibitem{Cvetic_Youm}M. Cvetic, and D. Youm, 
Nucl. Phys. B 453 (1995) 259, hep-th/9505045.
\bibitem{Cvetic_Tseytlin}M. Cvetic, and A. A. Tseytlin, 
Phys. Lett. B 366 (1996) 95, hep-th/9510097.
\bibitem{BCWKLM}K. Behrndt, G. L. Cardoso, 
B. de Wit, R. Kallosh, D. L{\"u}st, and T. Mohaupt, \\
Nucl. Phys. B 488 (1997) 236, hep-th/9601015. 
\bibitem{tod} K. P. Tod, 
Phys. Lett. B 121 (1983) 241.
\bibitem{sabra} W. A. Sabra,
Mod. Phys. Lett. A 13 (1997) 239, hep-th/9708103.
\bibitem{GMT} J.P. Gauntlett, R.C. Myers, and P.K. Townsend, 
Class. Quantum Grav. 16 (1999) 1, hep-th/9809065.
\bibitem{GGHPR} J.P. Gauntlett, J. B. Gutowski, 
C. M. Hull, S. Pakis and S. Reall, \\
Class. Quant. Grav. 20 (2003) 4587, hep-th/0209114. 
\bibitem{gauntlett}
J. P. Gauntlett, 
Fortsch. Phys. 53 (2005) 468, hep-th/0501229.
%
\bibitem{Gauntlett_Pakis} J. P. Gauntlett and S. Pakis, 
Commun. Math. Phys. 247 (2004) 421, hep-th/0212008.
%
\bibitem{gutowski6d} J. B. Gutowski, D. Martelli and H. S. Reall,
Class. Quant. Grav. 20 (2003) 5049, hep-th/0306235.
\bibitem{Das}S. R. Das, hep-th/9602172.
\bibitem{Cvetic_Hull}M. Cvetic, and C. M. Hull,
Nucl. Phys. B 519 (1988) 141, hep-th/9709033.
\bibitem{BMPV} J. C. Breckenridge, R. C. Myers, A. W. Peet and C. Vafa, 
Phys. Lett. B 391 (1993) 93, hep-th/9602065.
\bibitem{herdeiro} C. A. R. Herdeiro, 
Nucl. Phys. B 582 (2000) 363, hep-th/0003063.
\bibitem{EEMR04}
H. Elvang, R. Emparan, D. Mateos, H.S. Reall,
Phys.Rev.Lett. 93 (2004) 211302, hep-th/0407065. 
\bibitem{EEMR05_1}
H. Elvang, R.Emparan, D. Mateos, H.S. Reall,
Phys.Rev. D 71 (2005) 024033, hep-th/0408120.
%
\bibitem{Bena_Kraus} I. Bena and P. Kraus, 
JHEP 0412 (2004) 070, hep-th/0408186. 
\bibitem{Kraus_Larsen} P. Kraus and F. Larsen, 
hep-th/0503219. 
\bibitem{EEMR05_2}
H. Elvang, R. Emparan, D. Mateos, H.S. Reall,
JHEP 0508 (2005) 042, hep-th/0504125.
\bibitem{gutowski} J. B. Gutowski, H. S. Reall, 
JHEP 0402 (2004) 006, hep-th/0401042. 
\bibitem{Ohta} N. Ohta,
Phys. Lett. B 403 (1997) 218, hep-th/9702164.
\bibitem{OPS}N. Ohta, K. L. Panigrahi and Sanjay,
Nucl.Phys. B 674 (2003) 306, hep-th/0306186.
\bibitem{Miao_Ohta} Y.-G. Miao and N. Ohta,
Phys. Lett. B 594 (2004) 218, hep-th/0404082.
\bibitem{Brinkmann}H. W. Brinkmann, 
Proc. Natl. Acad. Sci. U.S. 9 (1923) 1. 
\bibitem{lunin} O. Lunin, J. Maldacena and L. Maoz,
hep-th/0212210.
\bibitem{Belinski_Zakharov1}V. A. Belinski and V. E. Zakharov, 
Sov. Phys. JETP 48 (1978) 985. 
\bibitem{Belinski_Zakharov2}V. A. Belinski and V. E. Zakharov, 
Sov. Phys. JETP 50 (1979) 1.
\bibitem{Mishima_Iguchi}T. Mishima and H. Iguchi, hep-th/0504018.
\bibitem{CJS}E. Cremmer, B. Julia and J. Sherk, 
Phys. Lett. B 76 (1978) 409. 
\bibitem{bertolini} M. Bertolini, P. Fr\`e and M. Trigiante, 
Class. Quant. Grav. 16 (1999) 1519, hep-th/9811251.

\end{thebibliography}
\end{document}